\documentclass[]{aastex61}
\pdfoutput=1 
\usepackage{amsmath,amstext}
\usepackage[T1]{fontenc}
\usepackage{apjfonts} 
\usepackage[figure,figure*]{hypcap}

\shorttitle{Co-moving $Gaia$-TGAS stars}
\shortauthors{Bochanski et al.}
\submitjournal{Astronomical Journal}
\accepted{8 Feb 2018}
\begin{document}

\title{Fundamental Properties of Co-Moving Stars observed by $Gaia$}

\correspondingauthor{John J. Bochanski}
\email{jbochanski@rider.edu}

\author[0000-0001-7096-425X]{John J. Bochanski}
\affiliation{Rider University, Department of Chemistry, Biochemistry and Physics, 2083 Lawrenceville Road,Lawrenceville, NJ 08648}

\author[0000-0001-6251-0573]{Jacqueline K. Faherty}
\affiliation{American Museum of Natural History, Department of Astrophysics, Central Park West at 79th Street, New York, NY 10034, USA}

\author[0000-0002-2592-9612]{Jonathan Gagn\'e}

\affiliation{Carnegie Institution of Washington DTM, 5241 Broad Branch Road NW, Washington, DC~20015, USA}
\affiliation{NASA Sagan Fellow}

\author{Olivia Nelson}
\affiliation{American Museum of Natural History, Department of Astrophysics, Central Park West at 79th Street, New York, NY 10034, USA}

\author{Kristina Coker}
\affiliation{American Museum of Natural History, Department of Astrophysics, Central Park West at 79th Street, New York, NY 10034, USA}

\author{Iliya Smithka}
\affiliation{American Museum of Natural History, Department of Astrophysics, Central Park West at 79th Street, New York, NY 10034, USA}

\author{Deion Desir}
\affiliation{American Museum of Natural History, Department of Astrophysics, Central Park West at 79th Street, New York, NY 10034, USA}

\author{Chelsea Vasquez}
\affiliation{American Museum of Natural History, Department of Astrophysics, Central Park West at 79th Street, New York, NY 10034, USA}

\begin{abstract}
We have estimated fundamental parameters for a sample of co-moving stars observed by $Gaia$ and identified by \cite{2017AJ....153..257O}.  We matched the $Gaia$ observations to the 2MASS and WISE catalogs and fit MIST isochrones to the data, deriving estimates of the mass, radius, $[Fe/H]$, age, distance and extinction to 9,754 stars in the original sample of 10,606 stars.  We verify these estimates by comparing our new results to previous analyses of nearby stars, examining fiducial cluster properties, and estimating the power-law slope of the local present-day mass function.  A comparison to previous studies suggests that our mass estimates are robust, while metallicity and age estimates are increasingly uncertain.  We use our calculated masses to examine the properties of binaries in the sample, and show that separation of the pairs dominates the observed binding energies and expected lifetimes.   
\end{abstract}

\textit

\section{Introduction}\label{sec:intro}
Stars with similar space motions, also known as co-moving stars, are unique testbeds for stellar and Galactic investigations.  They encompass a variety of separations, from $\sim 1$ AU, up to the widest separations observed \citep[$\sim$ 10\,pc; ][]{2010ApJS..190....1R,2017AJ....153..257O}.  They can be used as probes of star formation \cite[i.e.,][]{2016A&A...590A..13E}, planetary system survival \citep{2013Natur.493..381K,2014ApJ...782...60K} and Galactic dynamics \citep{2010MNRAS.401..977J}.  Widely separated pairs are of particular utility, as they are sensitive to the mass spectrum of large perturbers in the Milky Way, including giant molecular clouds and black holes \citep{1985ApJ...290...15B,1986Icar...65...27W,1987ApJ...312..367W}.  They are also sensitive to the overall mass distribution since they are easily disrupted by Galactic tidal forces \citep[e.g.,][]{1976iecg.book.....O,2002MNRAS.329..897H,2010MNRAS.401..977J}.  When widely separated pairs of stars are found, the individual stellar members serve to form a cluster of stars with $N=2$, making them benchmarks for calibrating age and metallicity relations \citep[i.e.,][]{2012AJ....143...67D,2012ApJ...748...93R}.  

Modern surveys have identified thousands of co-moving binaries within a few kpc of the Sun \citep{2010AJ....139.2566D,2015AJ....150...57D,2017AJ....153..259O, 2017AJ....153..257O, 2017arXiv170407829A}.  These survey studies have identified a new population of stars with separations of $<1$ to 10\,pc ($10^2$--$10^6$\,AU) easily some of the widest pairs known.  These pairs were identified using parallaxes and proper motions from $Gaia$.  For the comparison of true 3D velocity vectors, radial velocities (RVs) are required \citep{2017arXiv170407829A,2017arXiv170903532P}. At the largest separations, the false positive rate grows, reaching $\sim 60\%$, making RVs critical for identifying true companions \citep{2017arXiv170903532P}. With the approach of $Gaia$ data release 2, which includes the RVs of millions of bright stars ($G < 12$), many more co-moving stars should be discovered.

Despite these advances, the fundamental properties of the widest pairs are not well constrained.  First, binary interactions, moving groups and other phase-space structure can produce stars with similar motions that may not have begun their existence as bound companions, breaking the common assumption of co-eval and co-metallicity.  Next, pair-identifying algorithms can fracture larger ensembles of co-moving stars, shredding moving groups into isolated pairs.  Algorithms designed to work in observable space (i.e., R.A., Dec., proper motion) are more prone to this issue than those working in $XYZUVW$ space.  Finally, since the identification is mostly derived from the astrometric properties of the stars, the fundamental properties of the stars themselves, masses, ages and metallicities, have not been characterized. 

In this paper, we explore the fundamental properties (mass, radius, [Fe/H], age, and extinction) of thousands of widely separated pairs.  We derive these properties from an ensemble of survey observations, along with isochronal fits to each star. In \autoref{sec:obs}, we describe the $Gaia$ observations and the catalog of \cite{2017AJ....153..257O}, which we use for this analysis.  The \cite{2017AJ....153..257O} catalog was recently reorganized and re-analyzed by \cite{faherty2017}, which we adopt for this analysis.  We supplement the $Gaia$ observations with archival photometry from ground and space-based telescopes.  In \autoref{sec:analysis}, we provide estimates of fundamental parameters for all stars in the catalog, including mass, age, and metallicity.  Our estimates are verified by comparisons to prior studies and known fiducial clusters, including the Pleiades and Hyades. In \autoref{sec:discussion}, we explore our results, including an analysis of the mass properties of the binary systems.  Finally, our conclusions are presented in \autoref{sec:conclusions}.

\section{Observations}\label{sec:obs}
Below we describe the observations used in our analysis.  Our catalog incorporates $Gaia$ photometry and astrometry, as well as photometry obtained during the Two-Micron All-Sky Survey \citep[2MASS;][]{2006AJ....131.1163S} and by the Wide-Field Infrared Sky Explorer satellite \citep[WISE;][]{2010AJ....140.1868W}.

\subsection{$Gaia$ Observations}
The $Gaia$ satellite \citep{2016A&A...595A...2G} was launched in December 2013 and will map the entire sky over 5 years, producing the largest and most precise astrometric catalog yet.  The final catalog is expected to contain the sky positions, proper motions and distances to $\sim$ 1 billion unique stars, with a typical parallax, $\varpi$, uncertainty of $\sigma_{\varpi} \sim$ 20 $\mu$as for a Sun-like star with $G$ = 15.  

The $Gaia$ satellite images the sky using two telescopes separated by a basic angle of 106.5$^{\circ}$ focusing light onto a focal plane of 106 CCDs \citep{2016A&A...595A...1G}.
The first data release from the $Gaia$ team included proper motions and parallaxes for about 2.0 million nearby bright stars observed by Tycho-2 and Hipparcos.  The Tycho-$Gaia$ Astrometric Solution catalog \cite[TGAS;][]{1997ESASP1200.....E,2000A&A...355L..27H,2007A&A...474..653V,2015A&A...574A.115M,2016A&A...595A...4L} contains mostly bright stars, with 90\% of the catalog having $G < 12.05$.  The median uncertainty in parallax and position is 0.32 mas, with proper motion uncertainties of 1.32 mas yr$^{-1}$ \citep{2016A&A...595A...4L}.  \cite{2017MNRAS.470.1360B} has shown that the TGAS catalog is mostly complete to a distance of $\sim$ 200 pc for main-sequence spectral types A through K.

The \cite{2017AJ....153..257O} sample of co-moving stars, later examined and re-organized by \cite{faherty2017}, was drawn from the TGAS sample.  We summarize the basic sample construction here, and refer the reader to \cite{2017AJ....153..257O} and \cite{faherty2017} for details. The \cite{2017AJ....153..257O} sample first applied a global signal-to-noise cut on the parallaxes in the TGAS catalog, retaining 619,618 stars with $[S/N]_\varpi$ > 8.  Next, they searched for stars with similar space motions and 3D separations $< 10$ pc.  They identified 271,232 pairs of stars meeting these criteria, then applied a statistical selection, based on a fully marginalized likelihood, to identify the most likely pairs.  The final \cite{2017AJ....153..257O} catalog contained 10,606 individual stars organized into 4,236 unique groups with over 319 of those groups containing 3 or more stars (triples or higher order). \cite{faherty2017} re-analyzed the groups identified by \cite{2017AJ....153..257O} using the BANYAN~$\Sigma$ code \citep[][]{2018arXiv180109051G,2017ApJS..228...18G,2017ApJ...841L...1G} as well as a literature search and found many of the hierarchical groups were parts of known clusters (i.e., the Pleiades, $\alpha$~Per, and the Hyades clusters) and nearby moving groups and associations (e.g., Lower Centaurus Crux, Upper Centaurus Lupus). However known stellar members of associations were also broken up across several unique \cite{2017AJ....153..257O} groups. For example, \cite{faherty2017} recorded Hyades members found within 8 groups identified by \cite{2017AJ....153..257O} and members of the Lower Centaurus Crux association were found in 26 different groups.  We refer the reader to \cite{faherty2017} for a complete discussion of the re-organized sample.

\subsection{Cross Matching with 2MASS and WISE}
In order to compare observed data on our pairs to model isochrones, we supplemented $Gaia$ photometry with near-infrared (NIR) data from 2MASS and mid infrared data (MIR) from the WISE mission.

The 2MASS project employed two identical 1.3m telescopes in the Northern and Southern hemispheres to systematically map the night sky in the $J$ (1.1\,$\mu$m), $H$ (1.8\,$\mu$m) and $K_s$ (2.2\,$\mu$m) bands.  The northern telescope was located at the Whipple Observatory on Mount Hopkins in Arizona, USA, while the southern telescope was found at the Cerro-Tololo Inter-American Observatory at Cerro-Tololo, Chile. Over the course of 3 years, the 2MASS project recorded 24.5 TB of raw images, resulting in an all-sky catalog of over 470 million objects, which are mostly point sources.  The point-source catalog (PSC) is complete to $J<15.8, H<15.1, K_s < 14.3$, when confusion is unimportant. Typical uncertainties on the photometric observations are $\sim 1-2\%$, while the astrometric uncertainties for the most of the PSC is 0.07-0.08 arcseconds.

The Wide-Field Infrared Survey Explorer \citep[WISE][]{2010AJ....140.1868W} satellite imaged the entire sky in four infrared bands, centered on 3.4, 4.6, 12, and 22\,$\mu$m, and named $W1$, $W2$, $W3$ and $W4$ respectively.  The mission surveyed the sky during 2010, covering most of the sky at least twice during that time.  The telescope recorded images of over 560 million objects during its primary mission, and was restarted to search for near--Earth asteroids as NEOWISE \citep{2011ApJ...731...53M}.  Combining both programs resulted in the AllWISE catalog \citep{2013yCat.2328....0C}, which contains 747,634,026 objects and is 95\% complete to $W1 < 17.1$, $W2 < 15.7$, $W3 < 11.5$ and $W4 < 7.7$.  Typical astrometric precision is 0.15 arcseconds. 

Using the Tool for OPerations on Catalogues And Tables \citep[TOPCAT][]{2005ASPC..347...29T}, we implemented a 1$\arcsec$ radial search between the positions in Gaia and those in the ALLWISE catalog and recovered $W1W2W3W4$ photometry.  ALLWISE also automatically identifies matches with the 2MASS point source catalog using a 2$\arcsec$ radius therefore we also recovered $JHK_{s}$ photometry with one TOPCAT query.   After implementing our match, 
598 Gaia positions lacked a WISE measurement, and an additional 13 lacked 2MASS photometry. Therefore the full sample we used in the isochrone analysis below contained 9,995 unique stars from the \cite{2017AJ....153..257O} catalog. 

\section{Analysis - Fundamental Parameter Estimation}\label{sec:analysis}

We used the \texttt{isochrone} python module \citep{2015ascl.soft03010M} to estimate the fundamental parameters (mass, age, radius, $[Fe/H]$, distance, and extinction) of each star in our sample.  The package uses the Mesa Isochrones and Stellar Track library \citep[MIST;][]{2016ApJS..222....8D,2016ApJ...823..102C,2011ApJS..192....3P,2013ApJS..208....4P,2015ApJS..220...15P} and computes the posterior probability of fundamental parameters given the data. The MIST isochrones span $[Fe/H]$ from -4.00 to -2.00 in 0.50 dex steps and from -2.00 to +0.50 in 0.25 dex steps, and $\log{\frac{\rm Age}{\rm Gyr}}$ from 5.0 to 10.3 in 0.05 dex steps.  The isochrones are available in many standard bandpass sets, including $Gaia$, 2MASS and WISE.\footnote{The isochrones are available at \url{http://waps.cfa.harvard.edu/MIST/}.} 

\begin{figure*}\label{fig:corner_example}
\gridline{\fig{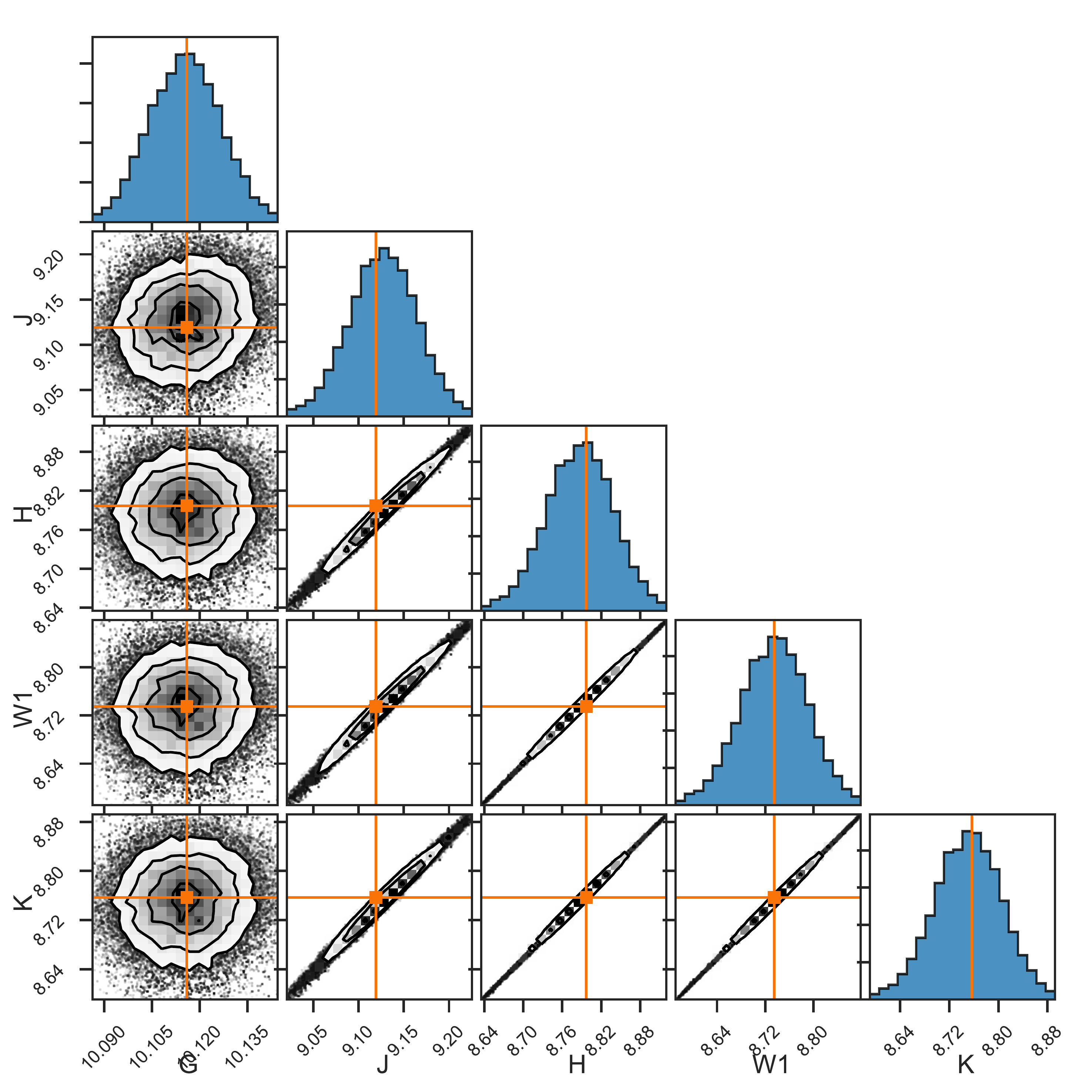}{0.55\textwidth}{}}
\gridline{\fig{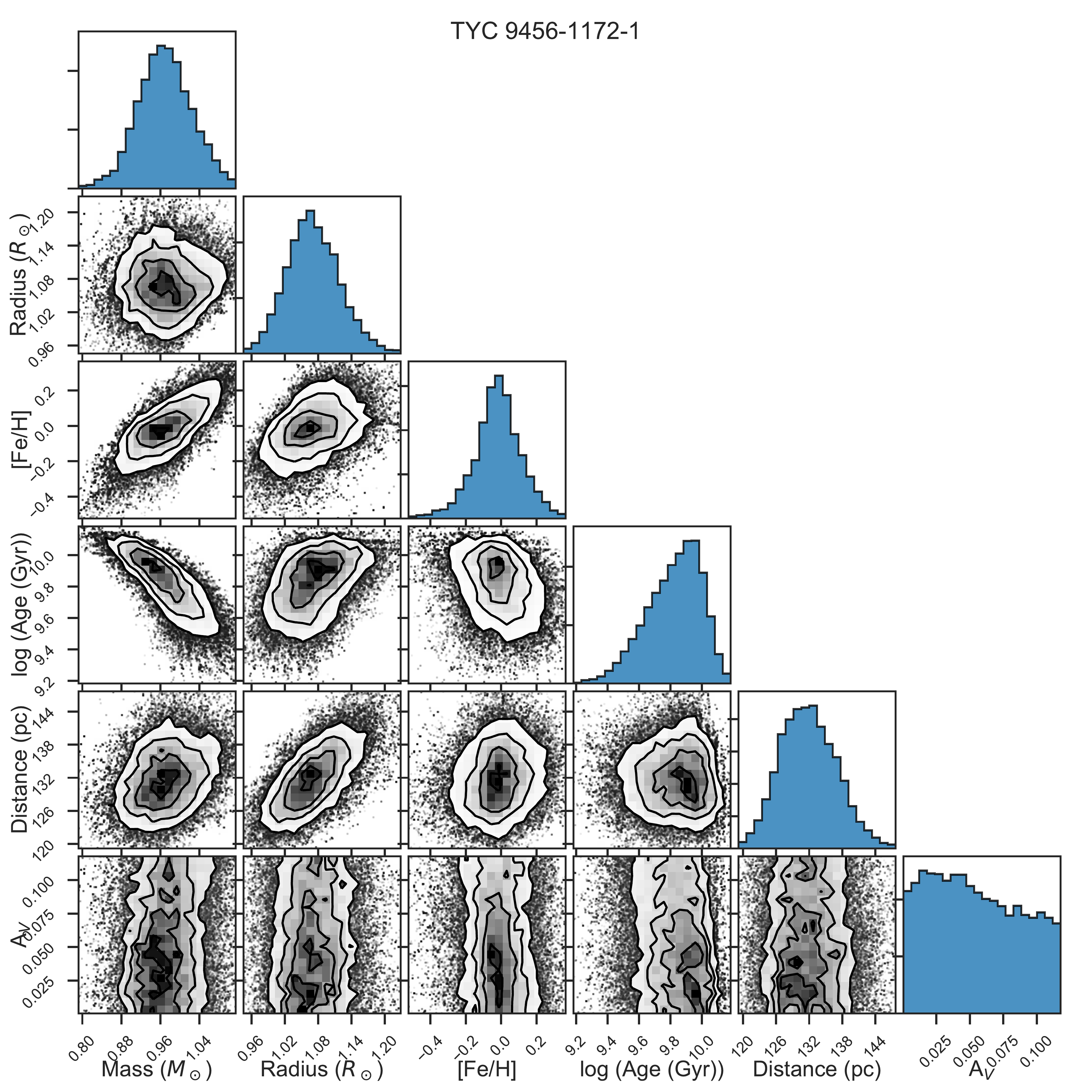}{0.55\textwidth}{}}
\caption{Upper Panel:  A corner plot \citep{2016JOSS.2016...24F} of posterior samples derived from our analysis for one example star, TYC 9456-1172-1.  The various panels are 2D projections along different axes of samples, along with histograms of the samples in $G,J,H,K$ and $W_1$.  The observed values are overplotted as blue vertical lines. Lower Panel:  A corner plot of the estimated fundamental parameters: mass, radius, [Fe/H], age, distance and $A_V$ for the star.}
\end{figure*}

\begin{figure*}\label{fig:corner_summary}
\plotone{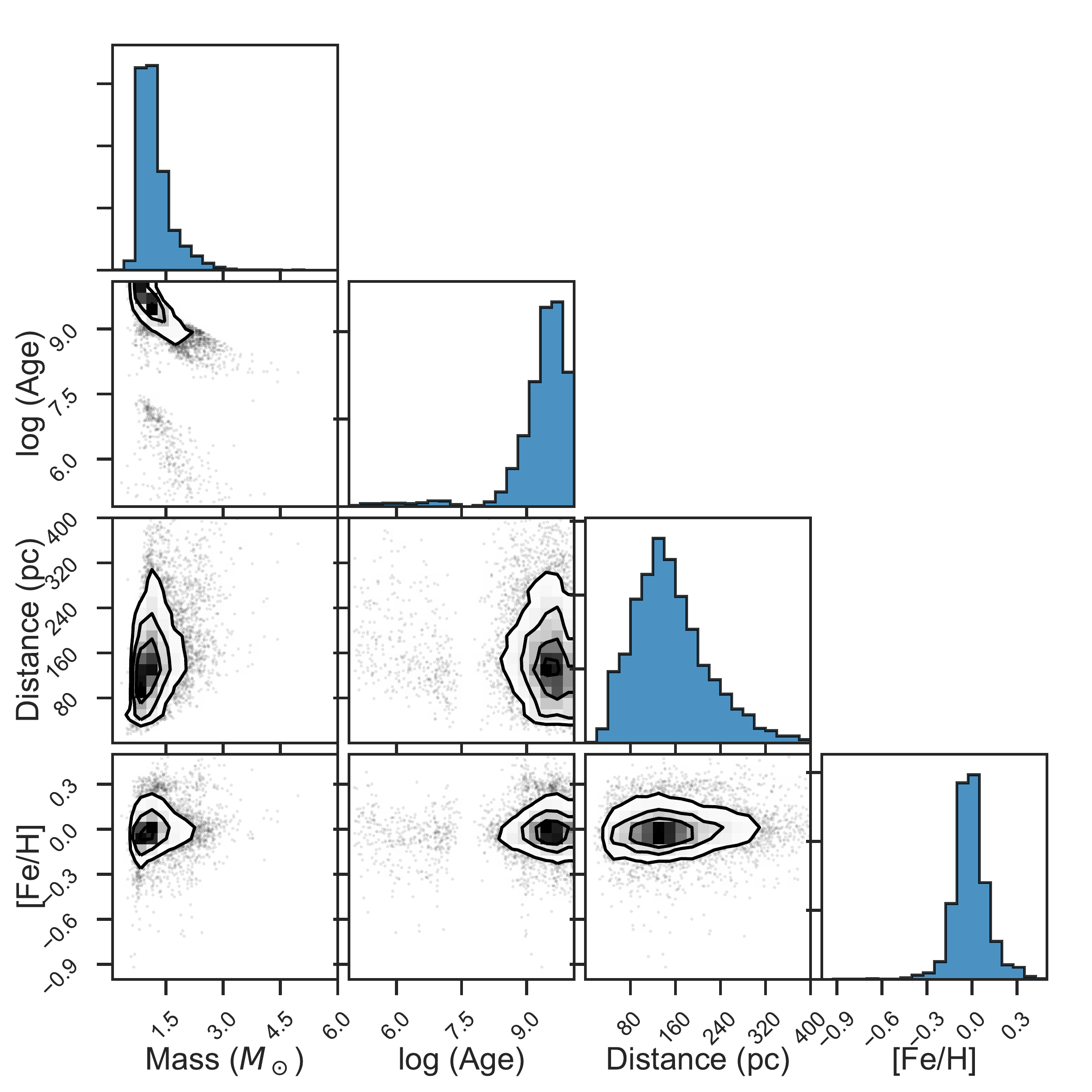}
\caption{Corner plot of fundamental parameter for our entire sample. Most stars in our sample are solar metallicity, with estimated ages of $\sim 10^9$ years. There are no obvious correlations between mass and metallicity or age, which suggests that our mass estimates are robust.}
\end{figure*}

We computed posterior probabilities on mass, age, radius, $[Fe/H]$, distance, and extinction for the sample conditioned on the measurements of $\varpi, G, J, H, K$, and $W1$ and their uncertainties. These posteriors were calculated using the trilinear interpolation schemes within \texttt{isochrones} and assumed priors described in \cite{2015ascl.soft03010M}, including a distance prior from the parallax reported by $Gaia$ and a $[Fe/H]$ prior from metallicity estimates \citep{2016ApJ...817...49B} of nearby stars by \cite{2011A&A...530A.138C}.  The extinction prior for each star was bounded at its maximum value with an extinction estimate calculated from the reddening reported in the \cite{1998ApJ...500..525S}, using the re-calibrated values \citep{2011ApJ...737..103S} and the \cite{1999PASP..111...63F} reddening law.  Next, the posterior distributions were sampled using the MCMC ensemble sampler \texttt{emcee} \citep{2013PASP..125..306F}.  We initialized 500 walkers with a random initialization bounded by values described by \cite{2015ascl.soft03010M} and allowed them to explore the posterior probabilities for 100 iterations.  The sampler was then re-initialized as at the location with the highest likelihood, with a small Gaussian perturbation in all dimensions.  The walkers then ran for 700 steps, with a burn-in of 200 steps, and the last 500 steps were recorded. The fundamental parameters listed were obtained by taking the median posterior sample, along with the 15\% and 85\% percentile samples.  We compared the posterior samples of each parameter to their priors and observed differences, indicating the supplemental photometry aided in constraining the model parameters. An example of our posterior samples are shown in \autoref{fig:corner_example}. For 241 stars, the sampler was unable to converge upon a solution.  Therefore we report fundamental parameters for 9,754 of our input sample. All resultant fundamental parameters (mass, age, radius, $[Fe/H]$, distance, and extinction) are listed in \autoref{table:fundamental_param} and summarized in \autoref{fig:corner_summary}. 

\begin{deluxetable*}{lrrrrrrrr}\caption{Fundamental Parameters}\label{table:fundamental_param}
\tablehead{
\colhead{TGAS Source ID} &
\colhead{R.A. (deg)} &
\colhead{Dec. (deg)} &
\colhead{Mass ($M_\odot$)} &
\colhead{Radius ($R_\odot$)}&
\colhead{[Fe/H] (dex)}&
\colhead{log(age (yr))}&
\colhead{Distance (pc)}&
\colhead{$A_v$ (mag)}
}
\startdata
49809491645958528 &  59.4573 &  18.5622 & $0.81^{+0.06}_{-0.06}$ & $0.83^{+0.03}_{-0.04}$ & $-0.09^{+0.15}_{-0.16}$ & $9.97^{+0.34}_{-0.23}$ & $130.8^{+ 5.0}_{- 5.4}$ & $0.10^{+0.07}_{-0.11}$ \\66939848447027584 &  57.0704 &  25.2149 & $1.22^{+0.05}_{-0.05}$ & $1.53^{+0.09}_{-0.10}$ & $0.07^{+0.09}_{-0.11}$ & $9.58^{+0.10}_{-0.08}$ & $130.4^{+ 6.1}_{- 6.9}$ & $0.07^{+0.05}_{-0.05}$ \\50905051903831680 &  58.0034 &  19.5967 & $1.07^{+0.05}_{-0.04}$ & $1.10^{+0.04}_{-0.05}$ & $0.12^{+0.15}_{-0.11}$ & $9.56^{+0.24}_{-0.18}$ & $147.4^{+ 5.8}_{- 6.4}$ & $0.11^{+0.07}_{-0.10}$ \\51452746133437696 &  59.5072 &  20.6766 & $1.21^{+0.05}_{-0.06}$ & $1.30^{+0.08}_{-0.09}$ & $-0.07^{+0.11}_{-0.13}$ & $9.30^{+0.29}_{-0.19}$ & $131.5^{+ 8.2}_{- 8.7}$ & $0.13^{+0.07}_{-0.04}$ \\51619115986889472 &  58.3703 &  20.9072 & $0.79^{+0.03}_{-0.04}$ & $0.81^{+0.03}_{-0.03}$ & $0.11^{+0.11}_{-0.12}$ & $10.13^{+0.26}_{-0.13}$ & $132.2^{+ 4.2}_{- 4.4}$ & $0.09^{+0.06}_{-0.07}$ \\51674916201705344 &  58.8832 &  21.0793 & $0.84^{+0.06}_{-0.06}$ & $0.87^{+0.04}_{-0.04}$ & $-0.03^{+0.14}_{-0.13}$ & $9.97^{+0.35}_{-0.22}$ & $123.4^{+ 4.6}_{- 4.8}$ & $0.08^{+0.06}_{-0.07}$ \\51694741770737152 &  59.5902 &  21.2575 & $0.83^{+0.05}_{-0.05}$ & $0.85^{+0.03}_{-0.04}$ & $-0.03^{+0.14}_{-0.16}$ & $9.98^{+0.31}_{-0.20}$ & $138.0^{+ 5.3}_{- 5.6}$ & $0.08^{+0.06}_{-0.05}$ \\51742467447748224 &  58.6161 &  21.3895 & $0.92^{+0.05}_{-0.05}$ & $0.92^{+0.03}_{-0.04}$ & $-0.00^{+0.11}_{-0.09}$ & $9.75^{+0.42}_{-0.26}$ & $134.6^{+ 4.7}_{- 5.2}$ & $0.09^{+0.06}_{-0.06}$ \\51861420861864448 &  62.2309 &  20.3858 & $1.19^{+0.07}_{-0.07}$ & $1.30^{+0.07}_{-0.07}$ & $-0.04^{+0.14}_{-0.13}$ & $9.38^{+0.29}_{-0.20}$ & $126.1^{+ 6.0}_{- 6.4}$ & $0.13^{+0.09}_{-0.11}$ \\53783848223326976 &  60.9341 &  22.9441 & $1.16^{+0.09}_{-0.07}$ & $1.39^{+0.08}_{-0.09}$ & $0.09^{+0.16}_{-0.14}$ & $9.64^{+0.18}_{-0.15}$ & $137.3^{+ 6.3}_{- 7.1}$ & $0.15^{+0.09}_{-0.06}$ \\61519668439604992 &  52.8682 &  21.8217 & $1.29^{+0.01}_{-0.03}$ & $1.44^{+0.05}_{-0.04}$ & $-0.11^{+0.09}_{-0.06}$ & $9.27^{+0.14}_{-0.07}$ & $125.2^{+ 3.2}_{- 4.8}$ & $0.11^{+0.05}_{-0.05}$ \\62413983709539584 &  49.4574 &  22.8320 & $1.70^{+0.07}_{-0.10}$ & $2.17^{+0.12}_{-0.13}$ & $-0.03^{+0.13}_{-0.17}$ & $9.06^{+0.11}_{-0.08}$ & $129.3^{+ 4.9}_{- 5.3}$ & $0.10^{+0.07}_{-0.09}$ \\63052044051306112 &  56.2570 &  19.5592 & $1.21^{+0.08}_{-0.07}$ & $1.39^{+0.08}_{-0.09}$ & $-0.01^{+0.13}_{-0.15}$ & $9.47^{+0.27}_{-0.17}$ & $135.4^{+ 6.6}_{- 7.7}$ & $0.13^{+0.09}_{-0.08}$ \\63144712265008512 &  55.6245 &  20.1498 & $1.49^{+0.09}_{-0.09}$ & $1.58^{+0.10}_{-0.12}$ & $-0.04^{+0.16}_{-0.14}$ & $8.97^{+0.34}_{-0.20}$ & $130.2^{+ 7.3}_{- 8.8}$ & $0.11^{+0.08}_{-0.11}$ \\63289916519862656 &  56.5807 &  20.8796 & $0.82^{+0.06}_{-0.06}$ & $0.84^{+0.04}_{-0.04}$ & $-0.03^{+0.15}_{-0.16}$ & $10.02^{+0.32}_{-0.20}$ & $139.0^{+ 5.8}_{- 6.3}$ & $0.10^{+0.07}_{-0.06}$ \\63948214747182848 &  57.1640 &  21.9248 & $1.38^{+0.04}_{-0.04}$ & $1.70^{+0.10}_{-0.10}$ & $-0.01^{+0.04}_{-0.05}$ & $9.31^{+0.12}_{-0.08}$ & $117.6^{+ 4.2}_{- 4.6}$ & $0.08^{+0.06}_{-0.04}$ \\64053561704835584 &  57.4796 &  22.2440 & $2.38^{+0.06}_{-0.04}$ & $3.12^{+0.25}_{-0.21}$ & $-0.11^{+0.11}_{-0.14}$ & $8.72^{+0.02}_{-0.02}$ & $130.1^{+ 8.2}_{- 7.5}$ & $0.08^{+0.06}_{-0.07}$ \\64109911675780224 &  57.4092 &  22.5333 & $2.10^{+0.06}_{-0.08}$ & $1.91^{+0.12}_{-0.11}$ & $0.00^{+0.01}_{-0.01}$ & $8.48^{+0.20}_{-0.13}$ & $128.3^{+ 4.8}_{- 5.4}$ & $0.07^{+0.03}_{-0.05}$ \\64114241002810496 &  57.2970 &  22.6093 & $1.72^{+0.08}_{-0.08}$ & $1.83^{+0.12}_{-0.13}$ & $-0.01^{+0.11}_{-0.13}$ & $8.86^{+0.22}_{-0.14}$ & $130.6^{+ 6.2}_{- 6.6}$ & $0.10^{+0.07}_{-0.04}$ \\64172034082472448 &  57.5889 &  23.0962 & $0.84^{+0.05}_{-0.06}$ & $0.88^{+0.05}_{-0.07}$ & $-0.00^{+0.14}_{-0.15}$ & $10.04^{+0.37}_{-0.18}$ & $138.6^{+ 9.2}_{-11.2}$ & $0.10^{+0.07}_{-0.06}$ \\64313218247427456 &  54.8051 &  21.8431 & $2.03^{+0.08}_{-0.08}$ & $2.06^{+0.13}_{-0.13}$ & $0.03^{+0.06}_{-0.09}$ & $8.68^{+0.12}_{-0.10}$ & $140.3^{+ 6.8}_{- 6.8}$ & $0.13^{+0.08}_{-0.11}$ \\64317994252099840 &  55.6001 &  21.4733 & $1.11^{+0.04}_{-0.03}$ & $1.16^{+0.04}_{-0.04}$ & $-0.12^{+0.13}_{-0.11}$ & $9.41^{+0.13}_{-0.14}$ & $129.5^{+ 4.2}_{- 3.8}$ & $0.21^{+0.11}_{-0.11}$ \\64380632054140416 &  55.8798 &  22.1582 & $1.05^{+0.05}_{-0.04}$ & $1.07^{+0.04}_{-0.05}$ & $0.14^{+0.13}_{-0.11}$ & $9.63^{+0.24}_{-0.18}$ & $134.6^{+ 4.6}_{- 5.0}$ & $0.10^{+0.06}_{-0.09}$ \\64449729487990912 &  55.6002 &  22.4210 & $1.00^{+0.06}_{-0.06}$ & $1.08^{+0.05}_{-0.05}$ & $0.01^{+0.14}_{-0.14}$ & $9.78^{+0.22}_{-0.16}$ & $136.0^{+ 4.7}_{- 5.1}$ & $0.10^{+0.06}_{-0.07}$ \\64739244643463552 &  56.2456 &  22.0323 & $1.23^{+0.05}_{-0.05}$ & $1.36^{+0.07}_{-0.09}$ & $-0.01^{+0.07}_{-0.05}$ & $9.36^{+0.27}_{-0.16}$ & $142.3^{+ 6.1}_{- 7.5}$ & $0.09^{+0.06}_{-0.07}$ \\\enddata
\tablecomments{This stubtable is a preview of the entire sample, which will be available as a machine readable table (and at \url{https://github.com/jbochanski/gaia-wide-binaries/}.}
\end{deluxetable*}

\section{Discussion}\label{sec:discussion}
In the following section, we discuss the results of our analysis.  We begin with a validation of our analysis, by comparing our photometrically derived physical parameters to previous studies of the same stars.  This is followed by an analysis of members (bona fide and suspected) of the nearby clusters.  We also present color-magnitude diagrams with regard to fundamental parameters.  Next, we examine the distributions of masses, compositions and ages recovered by our analysis.  Finally, we discuss the mass ratio and binding energy distributions.  

\subsection{Validation}\label{sec:validation}
\subsubsection{Comparison to Geneva-Copenhagen Survey Members}\label{sec:GCS}
We compare our photometrically derived fundamental parameters to the catalog of \cite{2011A&A...530A.138C}, who re-analyzed the Geneva-Copenhagen Survey (GCS) of 16,682 FGK dwarfs with Stromgren photometry.  Their work re-calibrated the temperature scale, and resulted in mass, age and metallicity estimates of the GCS sample. The catalogs were matched on Hipparcos catalog number. This yielded 672 matches between our sample and the \cite{2011A&A...530A.138C} catalog.  

\begin{figure*}\label{fig:GCS}
\gridline{\fig{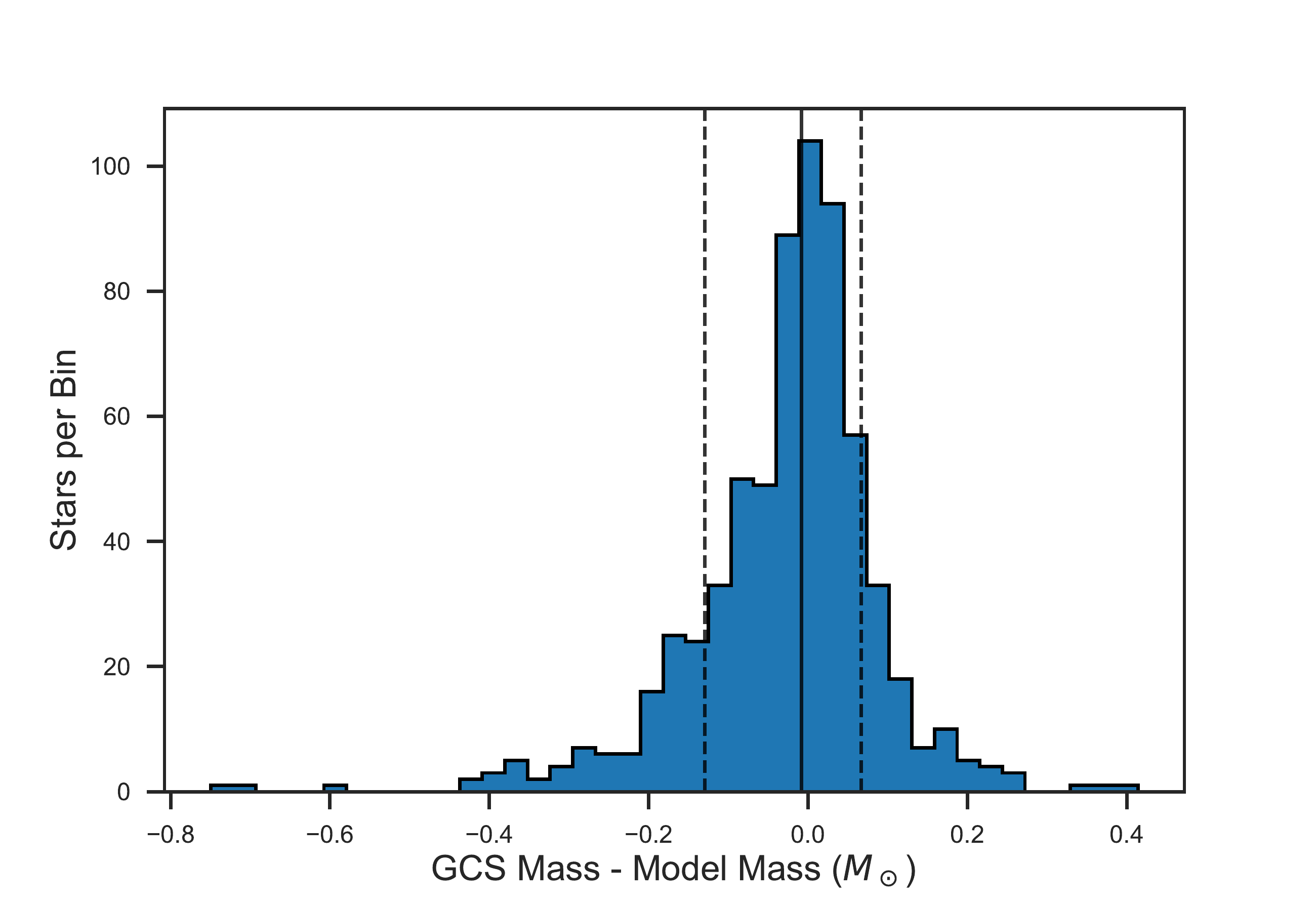}{0.5\textwidth}{}
			\fig{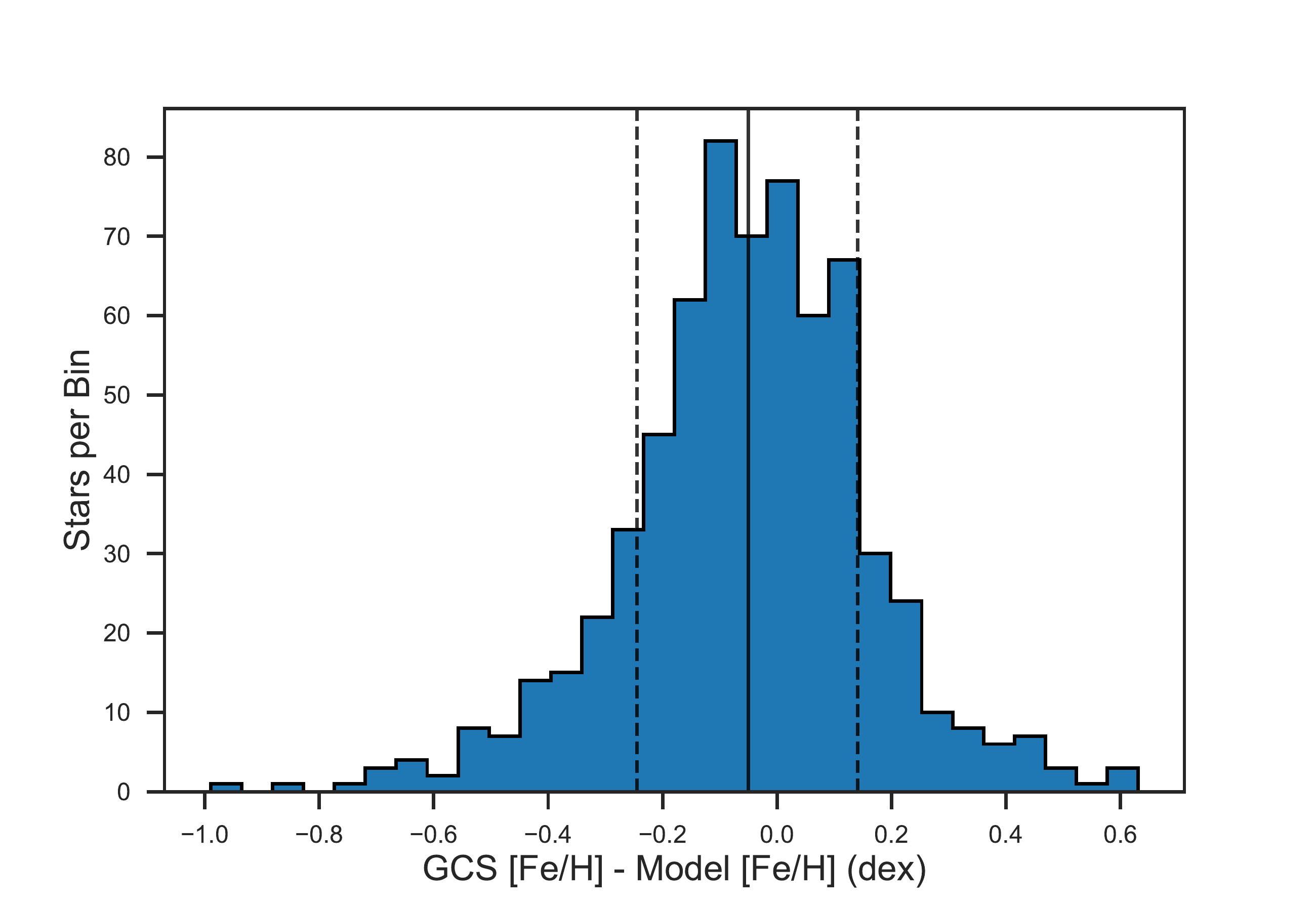}{0.5\textwidth}{}}
\gridline{\fig{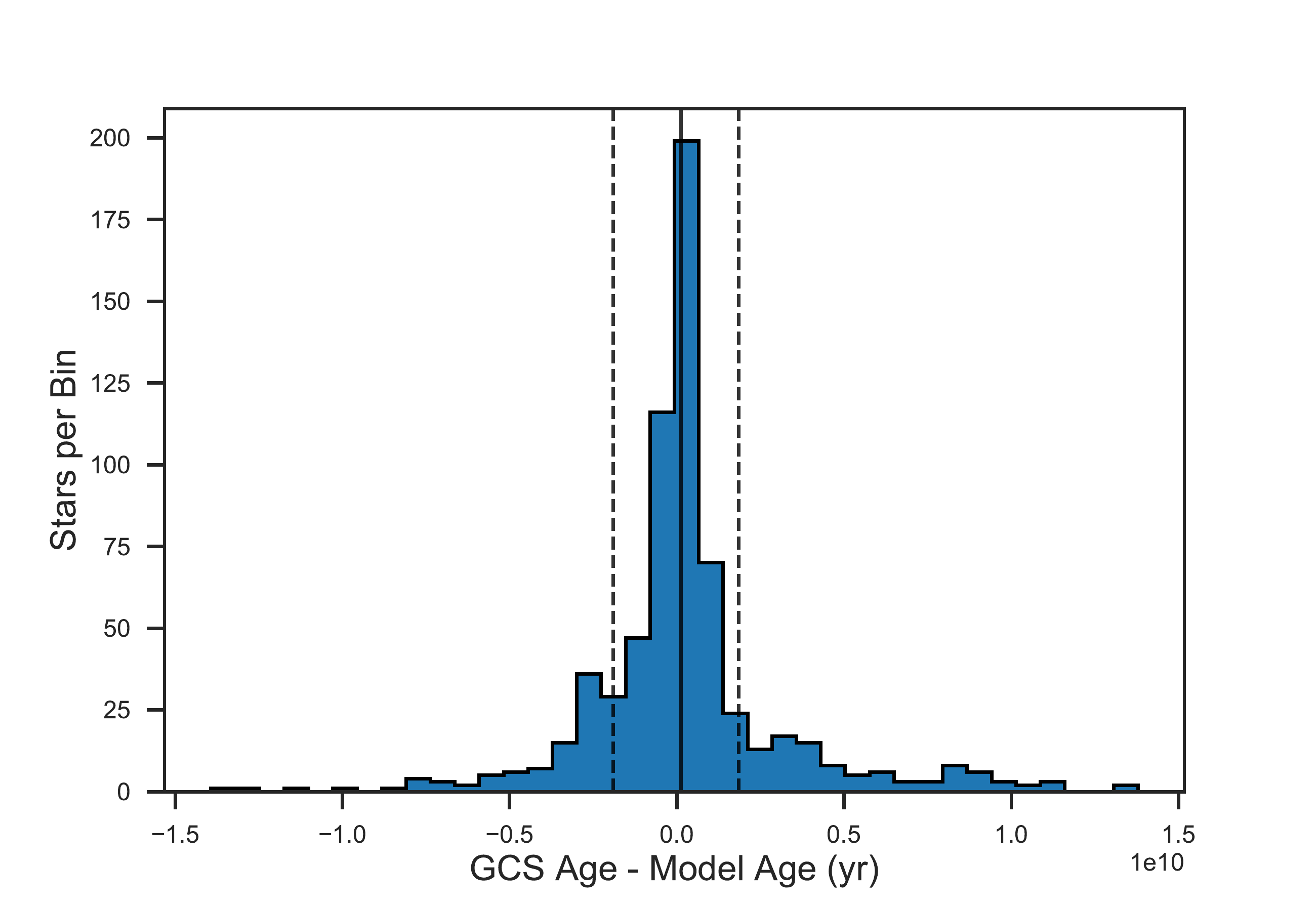}{0.5\textwidth}{}}
\caption{We compare the masses (upper left) , metallicities (upper right) and ages (lower) derived in our analysis, compared to the same properties derived by \cite{2011A&A...530A.138C} for the 672 stars in common between GCS and our sample.  We use the maximum likelihood values reported for the GCS stars using the Padova isochrones.   Overall, the best agreement is seen in the mass estimates of the two analyses.  We calculated the median and 15th and 85th percentiles (solid and dashed vertical lines) of the differences (GCS - this analysis).  They are -0.03 $^{+0.08}_{-0.12} M_\odot$ for mass, 0.13 $^{+1.73}_{-2.03} \times 10^9$ yr for age, and -0.05 $^{+0.19}_{-0.19}$ dex for metallicity.  }
\end{figure*}

In \autoref{fig:GCS}, we compare our fundamental parameter results to those from GCS for the stars in common between the two. Overall, the agreement in mass between the two catalogs is good.  The same agreement is not seen with age and metallicity, indicating that these parameters may be less certain.  We calculated the median and 15th and 85th percentiles of the differences between the two surveys (in the sense of GCS - this study).  They are -0.03 $^{+0.08}_{-0.12} M_\odot$ for mass (using the Padova isochrones), 0.13 $^{+1.73}_{-2.03} \times 10^9$ yr for age, and -0.05 $^{+0.19}_{-0.19}$ dex for metallicity. In each case, the median difference between the two samples is consistent with zero.  

\subsubsection{Comparing to Nearby Cluster and Moving Group Members}
\begin{figure*}\label{fig:cluster_comp}
\gridline{\fig{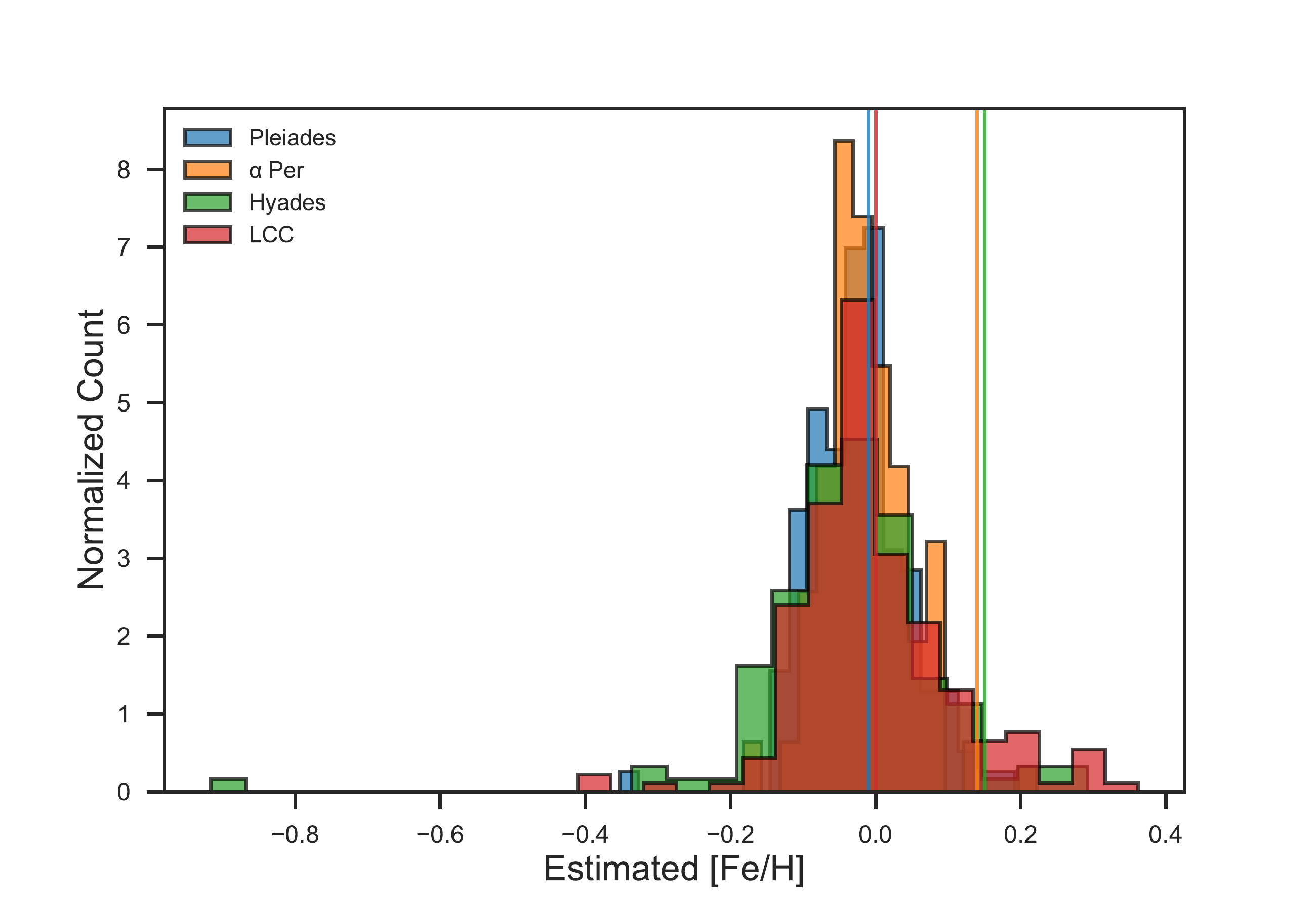}{0.5\textwidth}{}
		  \fig{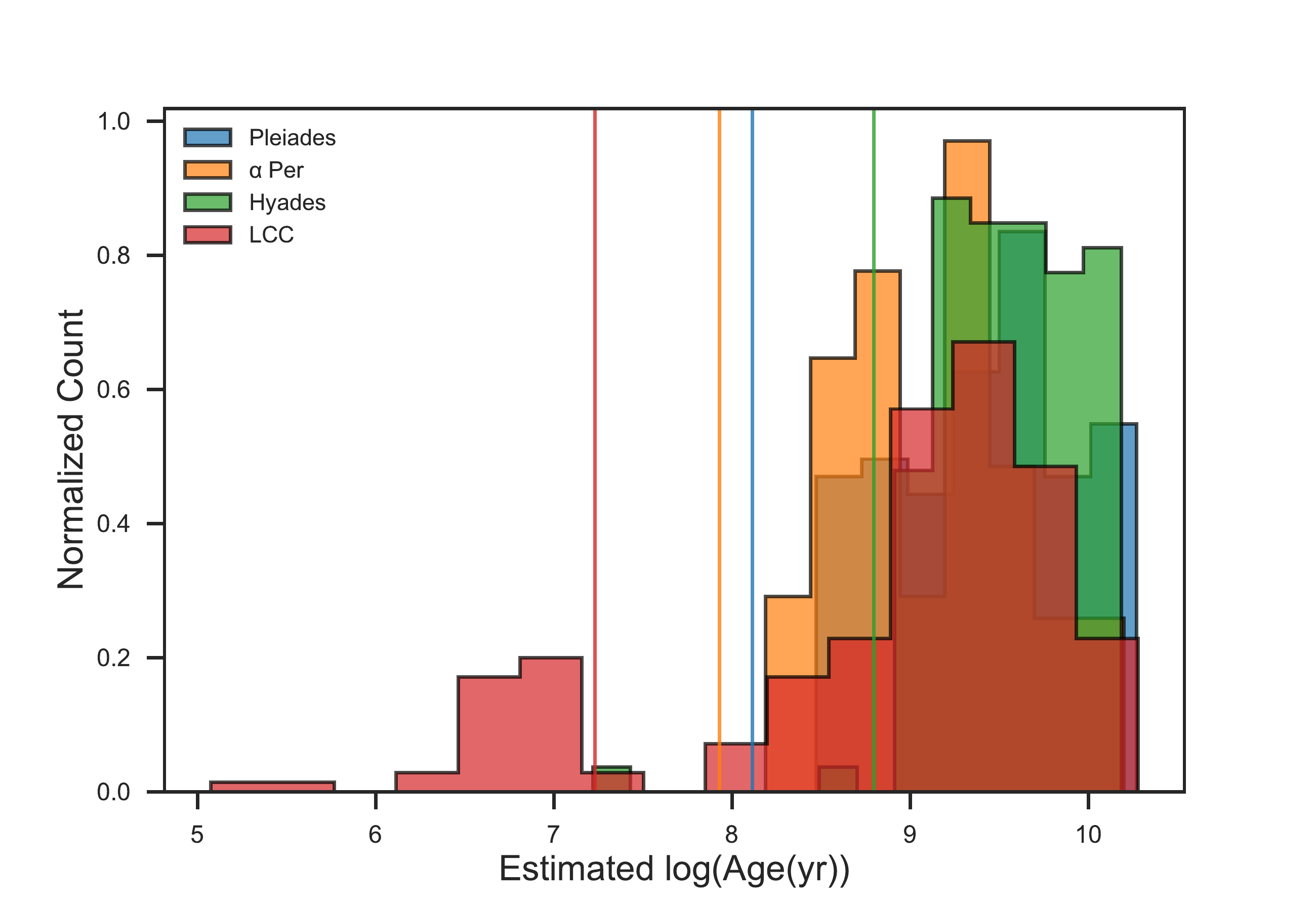}{0.5\textwidth}{}}
\gridline{\fig{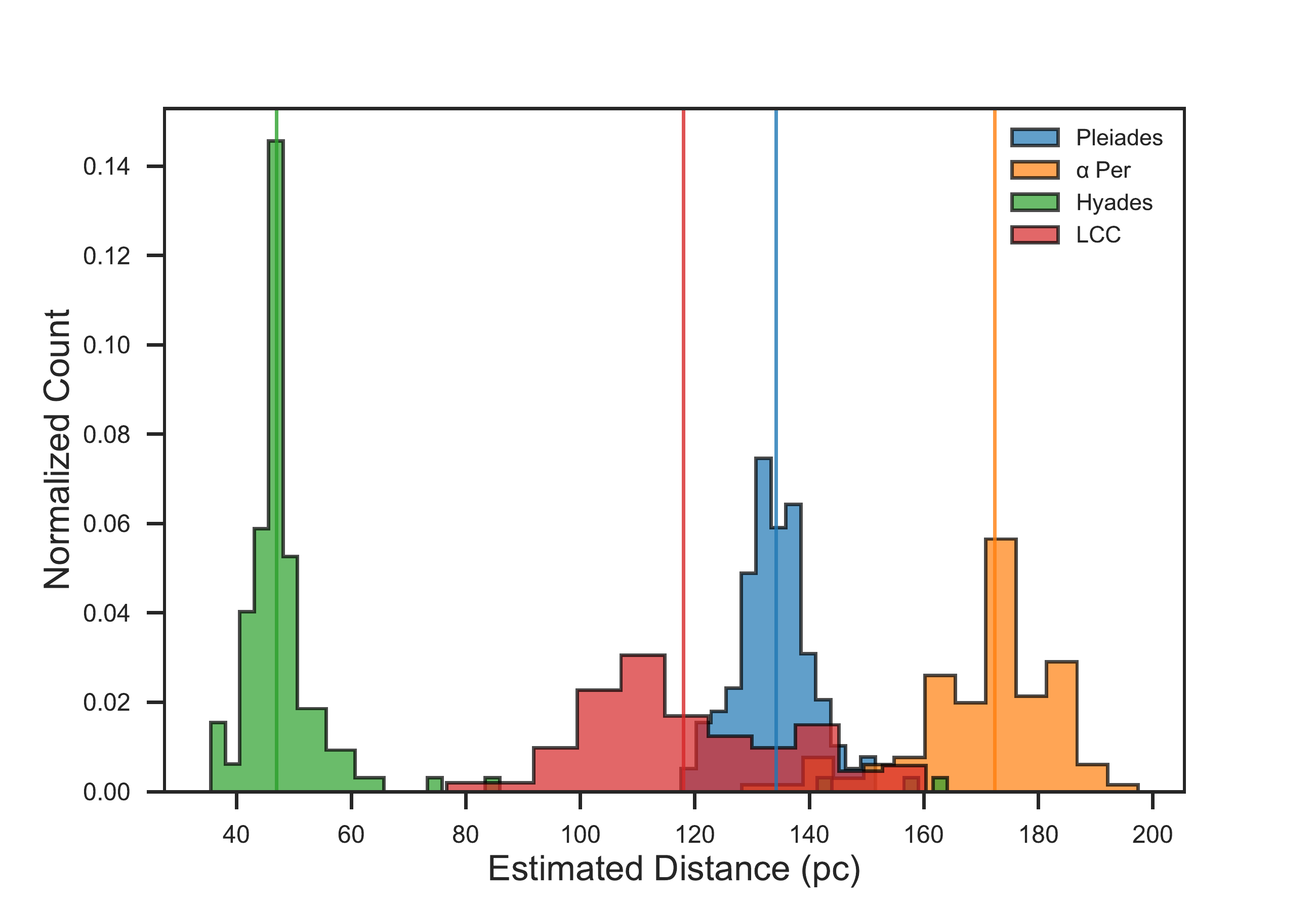}{0.5\textwidth}{}
		\fig{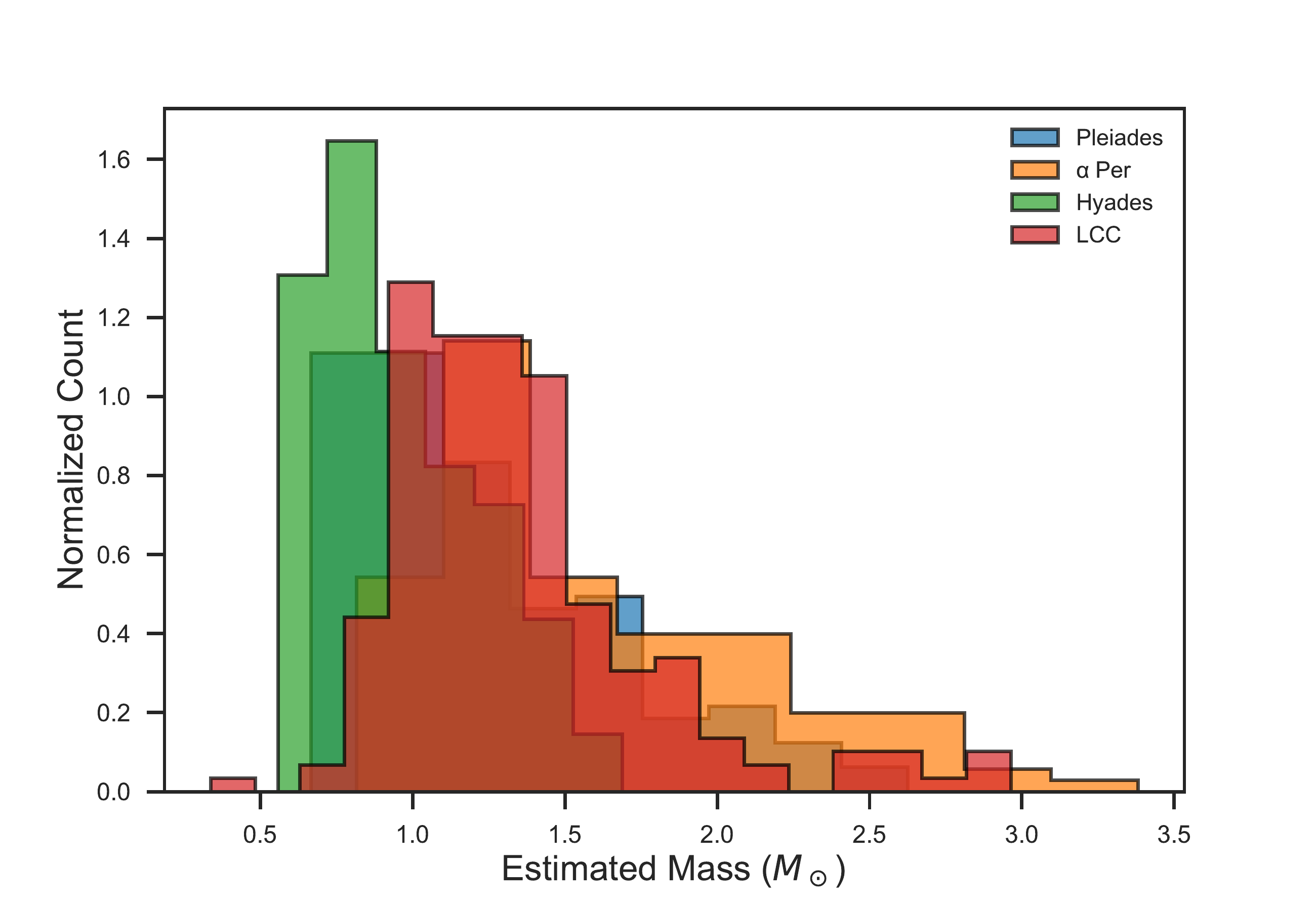}{0.5\textwidth}{}}
\caption{Histograms of fundamental parameters of identified members of the Pleiades (blue), $\alpha$~Per cluster(orange), Hyades (green) and Lower Centaurus-Crux moving group (red). Upper Left Panel: Most cluster members are found with metallicities close to solar.  Literature estimates for mean metallicities are over-plotted as vertical lines  Upper Right Panel: The ages of cluster members are shown, along with the mean ages of each cluster.  The best age agreement is with the oldest cluster, the Hyades.  Lower Left Panel: Distance estimates to cluster members, with average literature estimates over-plotted.  Lower Right Panel: Mass estimates of cluster members.  All four clusters show similar mass distributions.}
\end{figure*}

Next, we compared the derived fundamental properties to known clusters from \cite{2017AJ....153..257O} as identified in the reorganized catalog of \cite{faherty2017}.  The groups with the largest number of members were the Pleiades \citep{1921PASP...33..214C,1989ApJ...344L..21S}, $\alpha$~Per \citep{1974AJ.....79..687C}, and the Hyades open clusters \citep{1998A&A...331...81P}, and the Lower Centaurus Crux (LCC) group \citep{1946PGro...52....1B,1999AJ....117..354D}.   In \autoref{fig:cluster_comp} we show the derived fundamental parameters for the cluster members.  The metallicity distributions for the four associations are quite similar, containing mostly solar metallicity stars, in agreement with most spectroscopic results.  We over-plot literature estimates of the cluster metallicities as vertical lines.  These are -0.01 for the Pleiades, 0.15 for the Hyades, 0.14 for $\alpha$~Per, and 0.0 for the LCC \citep{2016A&A...585A.150N,2017AJ....153..128C}.  The age distribution is shown in the upper right, along with literature estimates of the age of each group.  The ages assumed are 130 Myr for the Pleiades \citep{2004ApJ...614..386B}, 85 Myr for $\alpha$~Per \citep{2004ApJ...614..386B}, 625 Myr for the Hyades \citep{1998A&A...331...81P} and 17 Myr for the LCC moving group \citep{2002AJ....124.1670M,2012ApJ...746..154P}.  Those values are over-plotted as vertical lines in the figure.  Overall, the agreement between our derived values and literature estimates (often derived spectroscopically) is marginal, but the enhancement of Myr stars found in LCC indicates that some age discrimination is possible with isochrone fitting. In the lower left panel, we compare the derived distance estimates (based on the isochronal fitting with a prior derived from $Gaia$ observations) to literature estimates of the mean distance for each cluster.  Due to the exquisite precision and accuracy of the $Gaia$ data, the agreement between our derived values and the literature values are good.  The adopted average distances for the Pleiades \citep{2016A&A...595A..59M}, $\alpha$~Per \citep{2009A&A...497..209V}, Hyades \citep{2007A&A...474..653V} and LCC \citep{1999AJ....117..354D} are 135 pc, 172 pc, 47 and 118 pc, respectively.  In lower right panel, we show the mass distributions of the four groups, which all share a similar slope.  The Pleiades and Hyades demonstrate a larger number of observed low-mass stars, with $\alpha$~Per containing a larger fraction of high mass stars.  However, since these are not complete surveys of the clusters, no strong statements can be made on intrinsic differences in the mass distributions.  Overall, we find good agreement between our analysis and literature values for the metallicity and distance estimates, with less agreement between age estimates.  This reflects the larger scatter seen in the age agreement in \autoref{sec:GCS} and the challenges in estimating ages from photometry.

\begin{figure*}\label{fig:clusterCMD}
\plotone{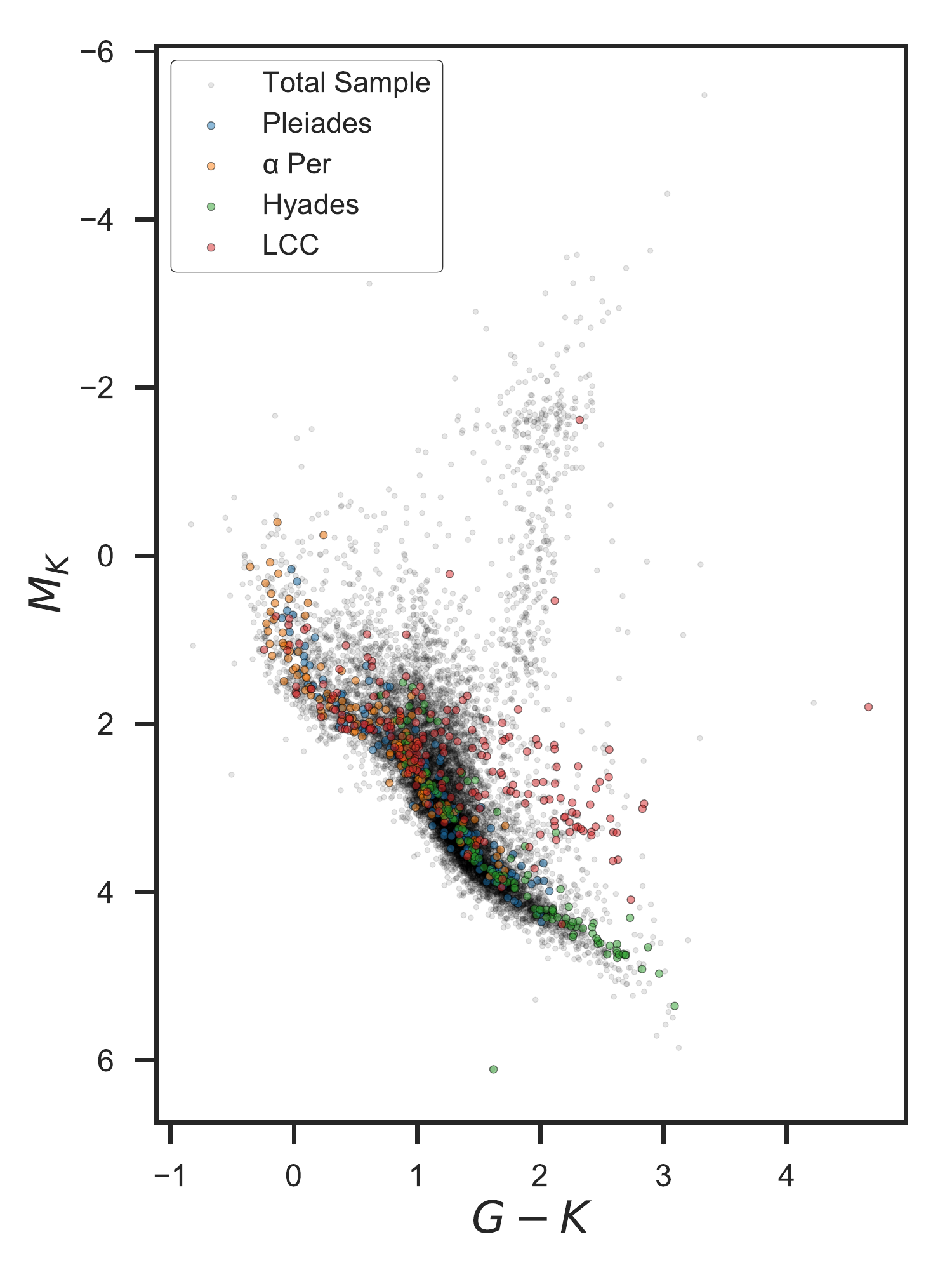}
\caption{Color Magnitude Diagram ($M_K$ vs. $G-K$) of our sample, members of the Pleiades (blue), $\alpha$~Per (orange), Hyades (green), and Lower Centaurus Crux (red) overplotted.  The Pleiades and Hyades contain the largest fraction of stars along the main sequence, while the LCC members show significant scatter.  This scatter is likely due to the young age of the LCC.}
\end{figure*}

In \autoref{fig:clusterCMD}, we plot the Gaia-2MASS color-magnitude diagram (CMD) of our total sample, along with members of the four clusters. The $G-K$ colors and $K$ magnitudes have not been corrected for extinction in any of the CMDS presented in this analysis.  The Pleiades, $\alpha$~Per and Hyades members occupy similar areas of color-magnitude space, due to their similar ages and compositions.  The LCC members demonstrate significant scatter. This is likely due to the the youth of the cluster, and the larger spread in distance, as seen in \autoref{fig:cluster_comp}.  For each cluster, stars are found above the main sequence.  For the older clusters, these are likely unresolved binaries, which are more common in wide binaries \citep{2010ApJ...720.1727L}, forming hierarchical multiple systems.  Note that the stars identified as cluster members are not necessarily wide binaries. 

\subsubsection{Present--Day Mass Function}
\begin{figure}\label{fig:mass}
\plotone{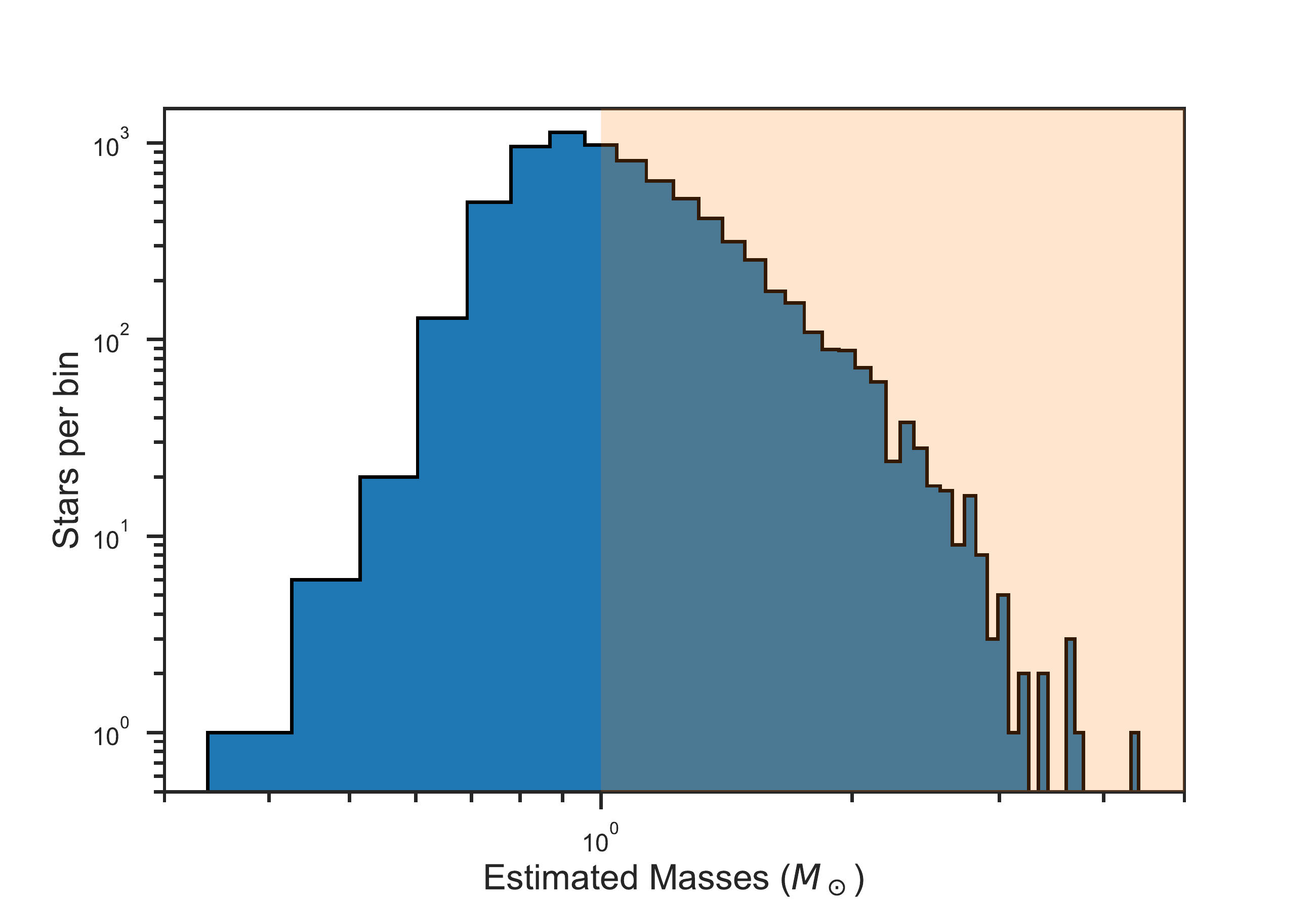}
\caption{Histogram of mass estimates derived in our analysis for stars with distances less than 200 pc, where TGAS is mostly complete for spectral types A through K.  We observe a peak at solar-mass stars, which should be complete in TGAS, and we note the falloff towards higher and lower masses.  The low-mass falloff is due to incompleteness at low masses, while the decline towards higher masses is due to the PDMF. We selected stars with estimated masses $> 1 M_\odot$ for our PDMF analysis, which is shaded above.}
\end{figure}

\begin{figure}\label{fig:mass_fit}
\plottwo{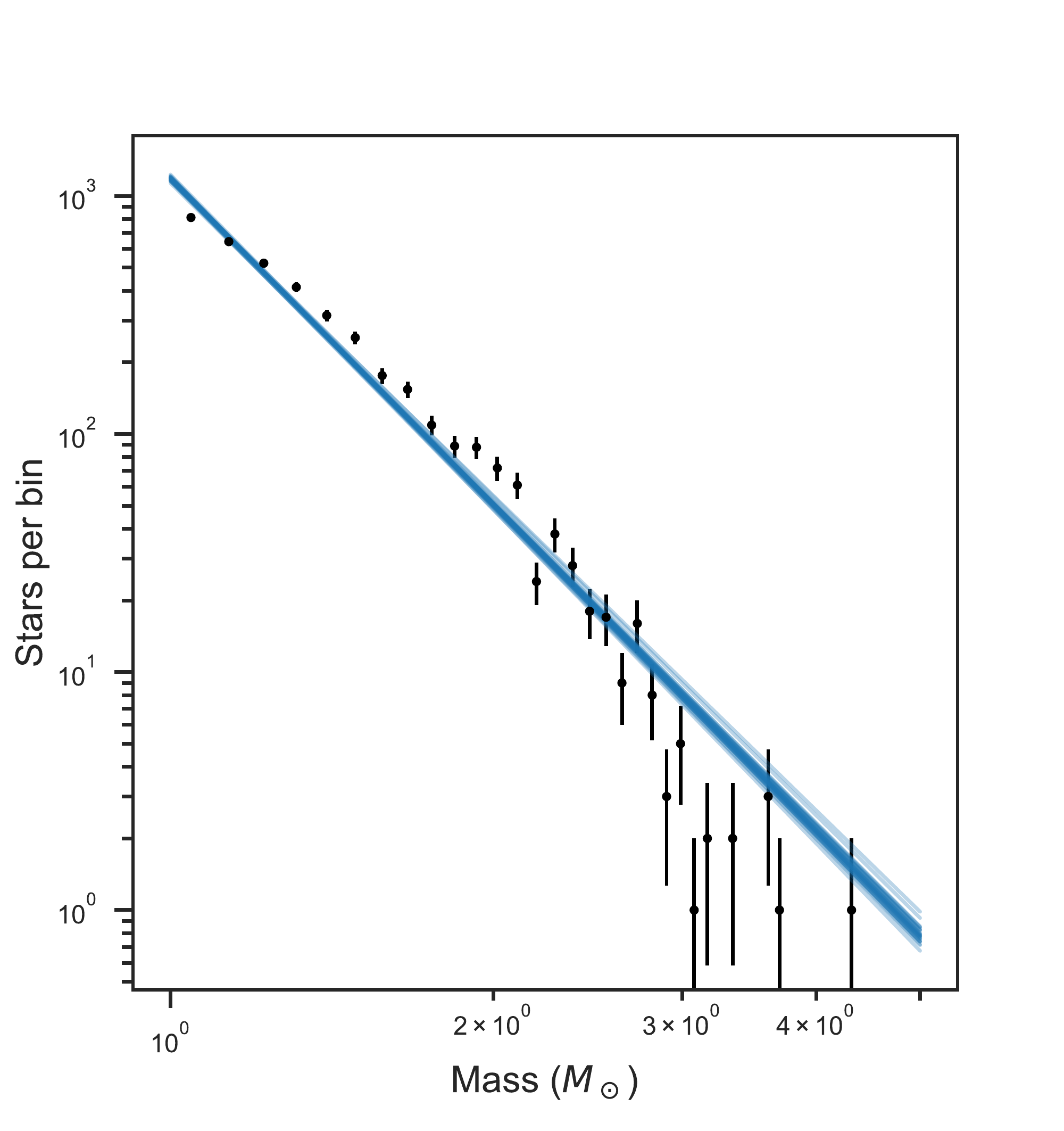}{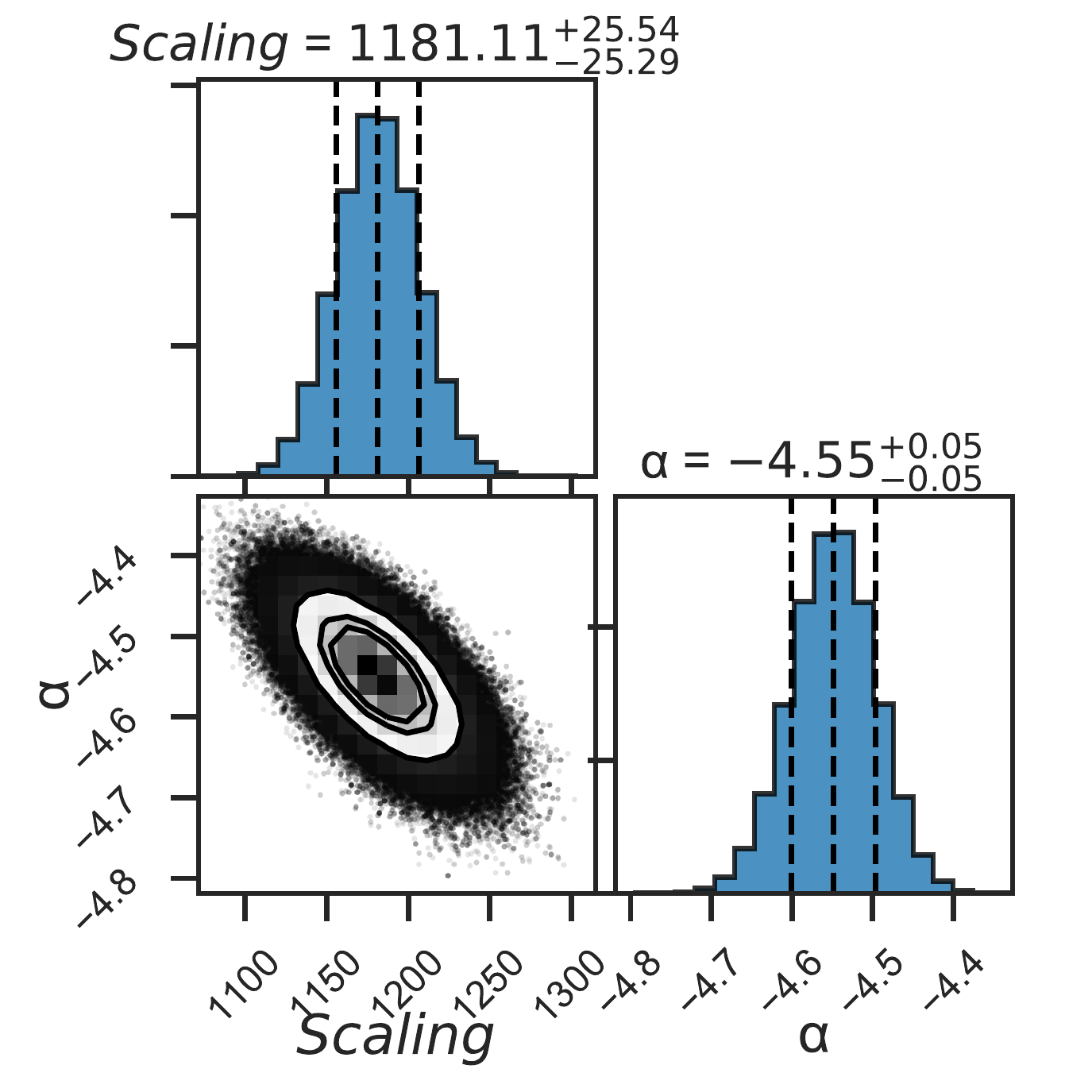}
\caption{Left Panel: Stars per mass bin, in log-log scale, for stars with $M > 1 M_\odot$ stars.  We sampled the posterior probabilities of a power-law fit to this distribution using \texttt{emcee}.  Right Panel:  The corner plot describing the fit to the mass distribution.  We do not attempt measure the overall scaling of the mass function (in terms of stars per pc$^3$), just the slope ($\alpha$) of the power-law fit.  The slope agrees well with current estimates of the PDMF \citep{2002AJ....124.2721R,2017MNRAS.470.1360B}.}
\end{figure}

We further scrutinized our mass estimates by determining the present--day mass function (PDMF).  The distribution of all masses can be found in \autoref{fig:corner_summary}.  For our PDMF measurement, we selected stars with masses $> 1 M_\odot$ within 200 pc, since TGAS is complete within 200 pc \cite{2017MNRAS.470.1360B} at these masses.  The histogram of mass estimates for stars within 200 pc are shown in \autoref{fig:mass}.  Bin sizes were chosen using Knuth's rule \citep{2006physics...5197K} as implemented in \texttt{astroML}\footnote{\url{http://www.astroml.org/}} \citep{2014ascl.soft07018V,astroML,astroMLText}.  Solar-mass stars are the most common member of our sample, with low-mass stars (0.2-1.0 $M_\odot$) as the next most common constituent.  The PDMF includes information on both the initial mass function \citep[IMF,][]{2010AJ....139.2679B} and the star formation rate.  In \autoref{fig:mass_fit}, we plot the number of stars per mass bin, in log-log scale, for stars with 1.0 $< M/M_\odot <$ 5.0 along with estimates of the posterior probability of a power-law fit.  The slope of the power-law, commonly given as $\alpha$, where $\alpha = -2.35$ is the slope measured by \cite{1955ApJ...121..161S}, was $\alpha = -4.55 \pm 0.05$, estimated by samples of the 16th, 50th, and 85th percentiles.  While our sample is not complete, it is well matched to the estimates of the PDMF for $M > 1 M_\odot$ by \cite{2002AJ....124.2721R}, $\alpha = -5.2 \pm 0.4$ and the recently derived PDMF from \cite{2017MNRAS.470.1360B} $\alpha = -4.7$, which used a different set of isochrones to estimate masses of TGAS stars.

\subsection{Color--Magnitude Diagrams}
\begin{figure*}\label{fig:CMDs}

\gridline{\fig{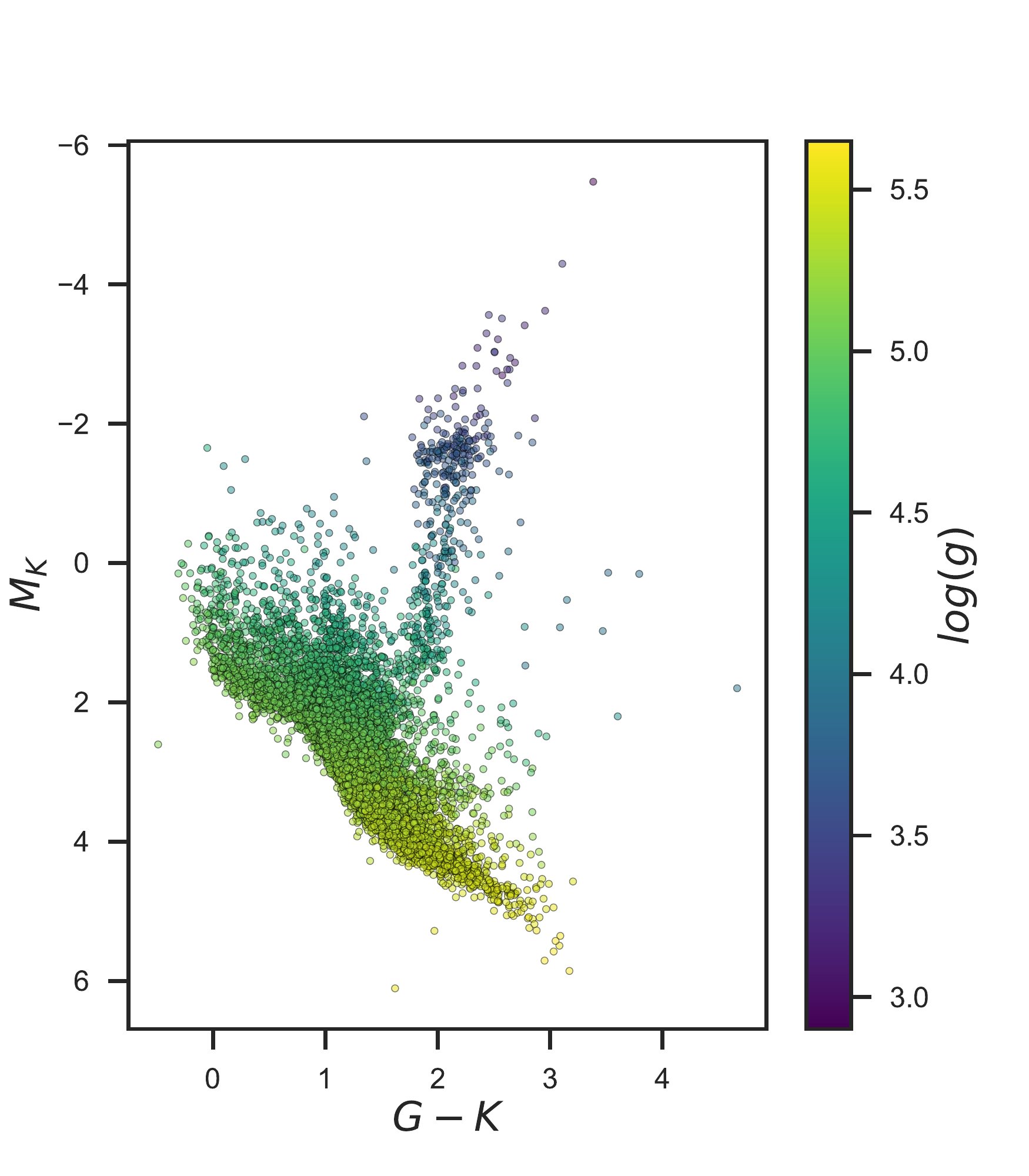}{0.5\textwidth}{(a) - log $g$}
          \fig{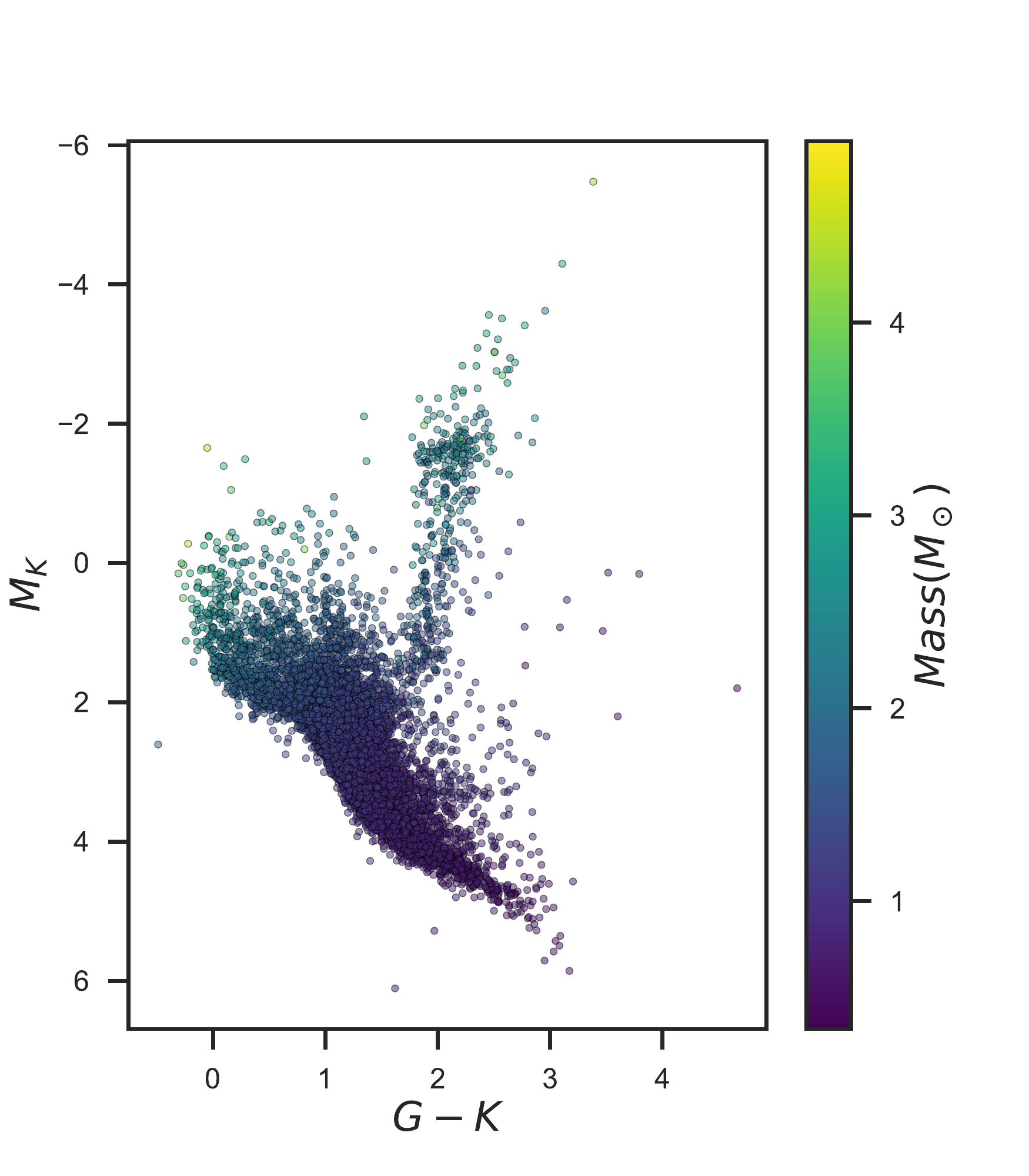}{0.5\textwidth}{(b) - Mass }}
\gridline{\fig{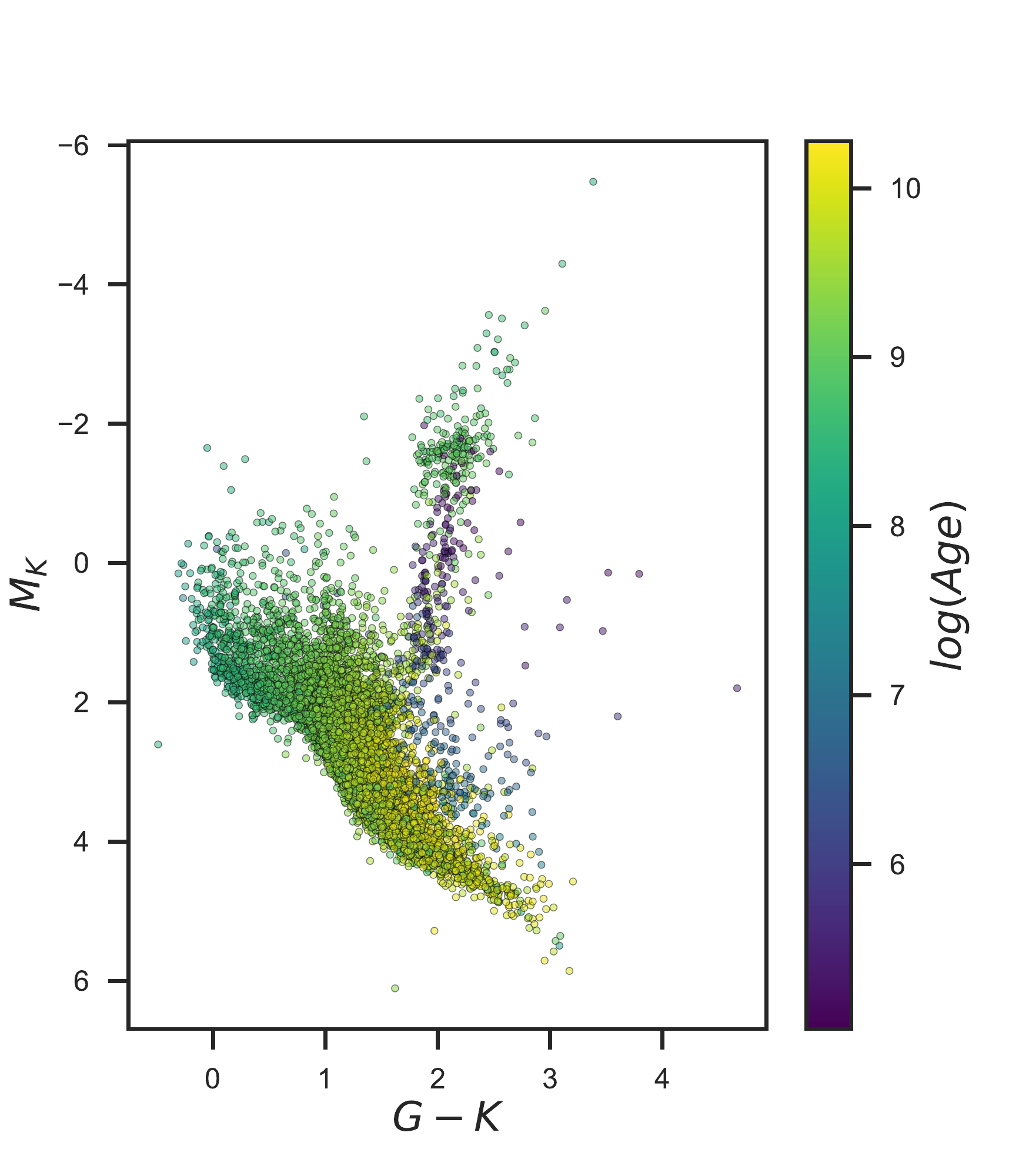}{0.5\textwidth}{(c) - Age }
          \fig{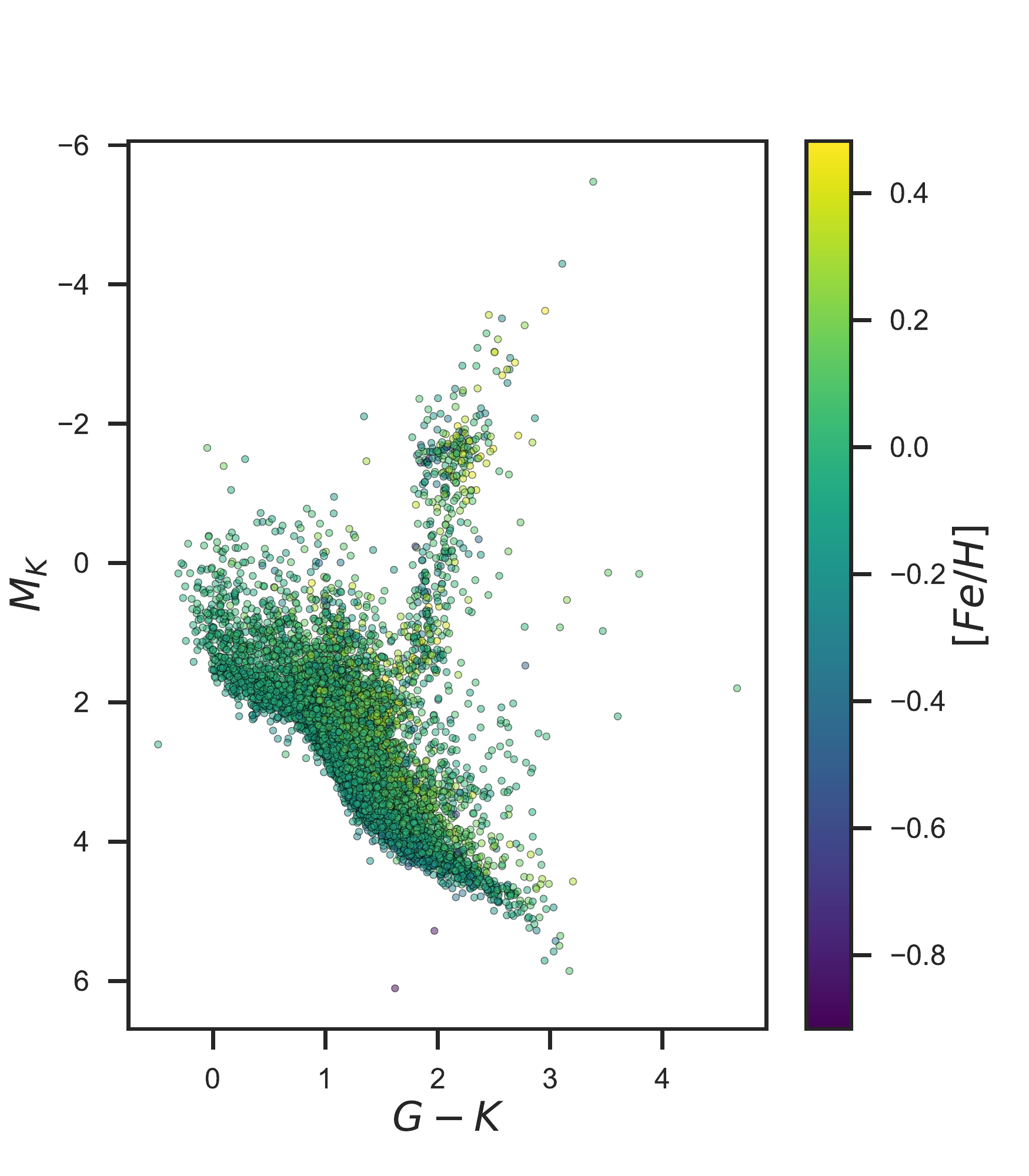}{0.5\textwidth}{(d) - [Fe/H]} }
\caption{Color--Magnitude Diagrams of our sample.  In each window, $M_K$ vs. $G-K$ is shown.  In panels a,b,c, and d each star is shaded by its log $g$, mass, age, and [Fe/H], respectively.}
\end{figure*}

In \autoref{fig:CMDs}, we plot color--magnitude diagrams of our sample in $M_K$ vs. $G-K$.  The upper left panel highlights the $N=2$ pairs, with a line connecting each component, after \cite{2017arXiv170903532P}.  The majority of the stars are found in pairs of main sequence stars, with a smaller subset containing a main sequence star with a  red giant branch stars.  

\begin{figure}
\plotone{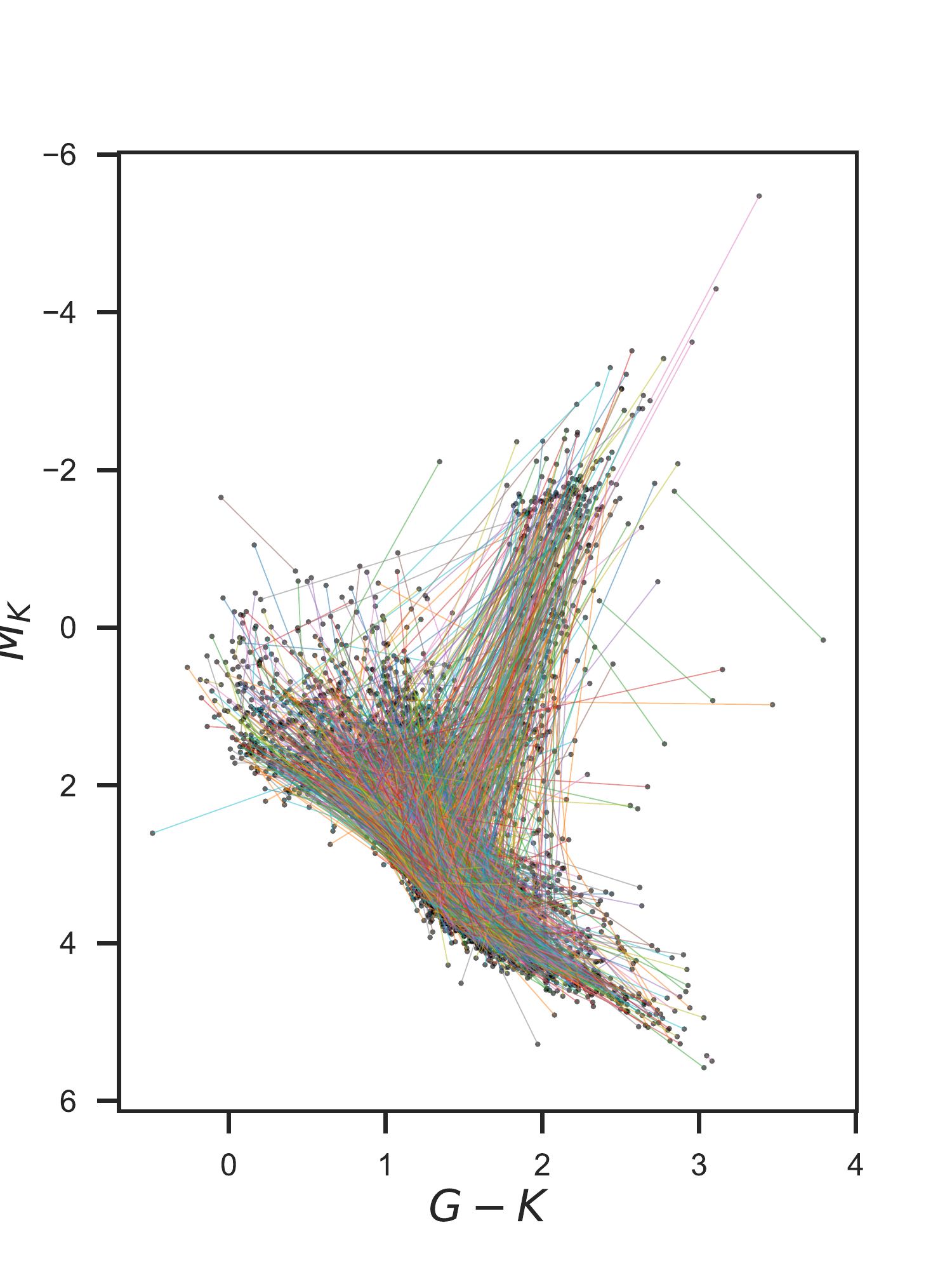}
\caption{The color magnitude ($M_K$ vs. $G-K$) for pairs in our sample, after \cite{2017arXiv170903532P}.  Each binary is connected by a colored line.  Most of the sample is composed of main sequence $+$ main sequence pairs, but some main sequence $+$ red giant branch stars are observed.}
\end{figure}

\subsection{Binary Properties}
We examined the properties of binaries in our sample.  Co-moving binaries (with $N=2$) are usually assumed to be coeval members with similar compositions, but recent results have shown that pairs may not always have the same metallicity \citep[i.e.,][]{2017arXiv170905344O}.  Below, we examine the differences in metallicity and age for pairs, and identify sets of "twin" stars in the sample.  We also highlight the properties of the ten most widely-separated pairs.  Finally, since our mass estimates are the most robust property measured, we examine the mass properties of the binaries, including their mass ratio and binding energy distributions, along with their expected dissipative lifetimes.

\subsubsection{Metallicity and Age Distributions}
\begin{figure}\label{fig:deltas}
\plottwo{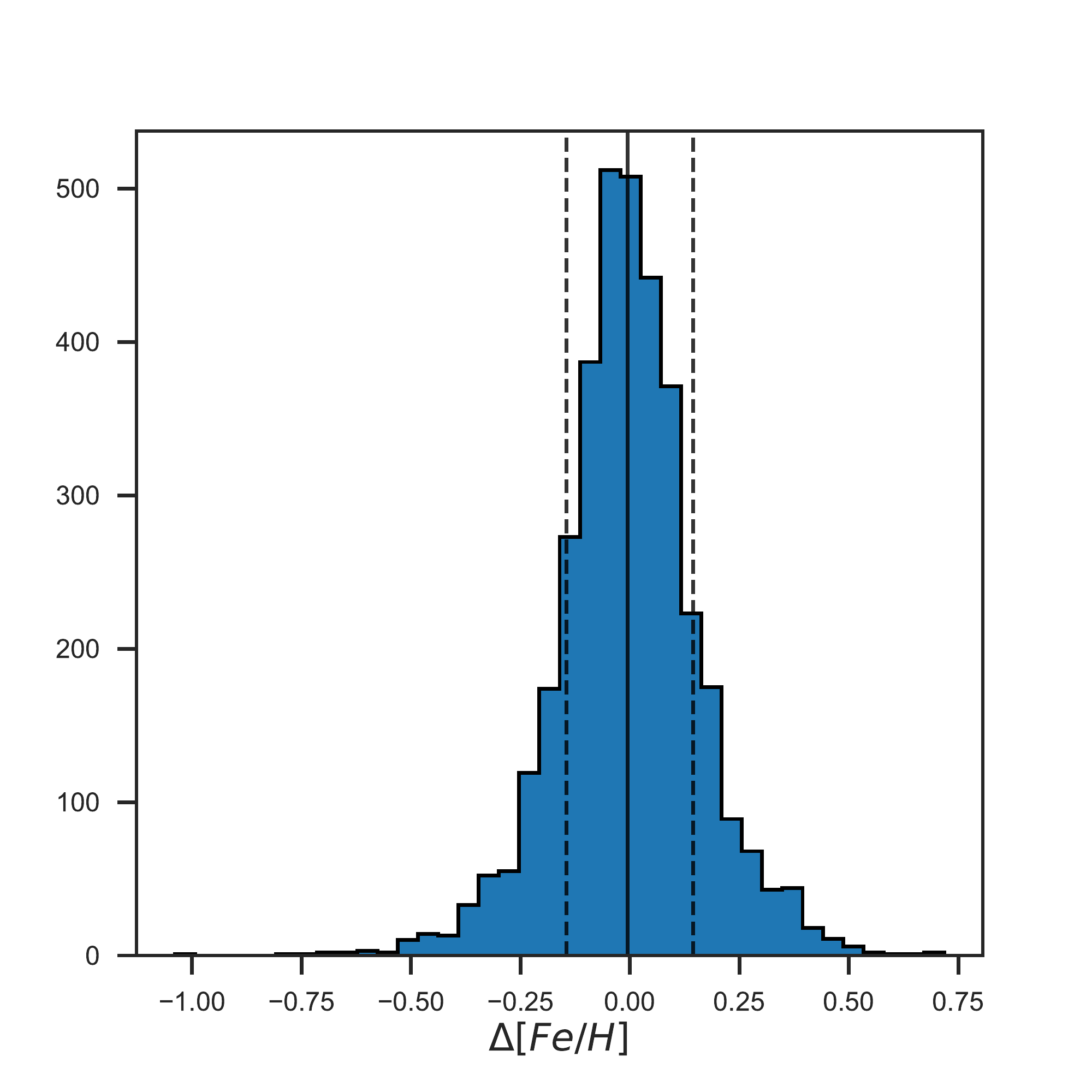}{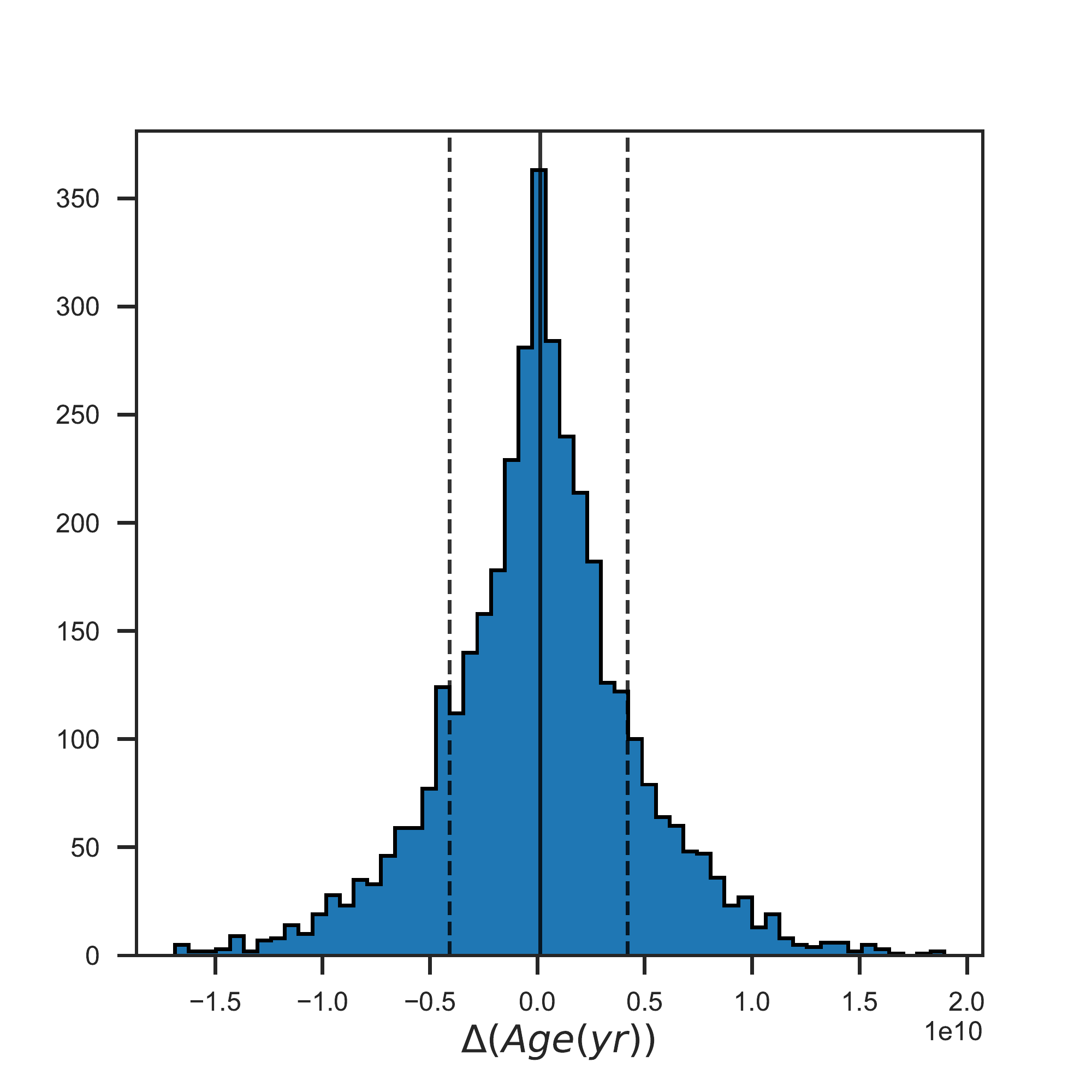}
\caption{Left panel:  Histogram of differences in [Fe/H] for binaries in our sample.  The distribution is peaked at 0 with most stars agreeing within $\pm 0.2$ dex (standard deviation = 0.16 dex), which matches our external precision.  Right Panel: Differences in ages between components.  There is less agreement in the age estimates (standard deviation = 4 Gyr), which is likely due to the difficulties in estimating age from photometry alone. In both panels, the median (solid vertical line) and 15th and 85th percentiles (dashed vertical lines) are overplotted.}
\end{figure}

As shown in \autoref{sec:validation}, metallicity and age have significant uncertainties when estimated using photometry alone.  In  \autoref{fig:deltas} we examine the difference in $[Fe/H]$ and log age (yr) for the sample.  In general the agreement in metallicity between members of binaries is good, with a broad peak centered on $\Delta_{[Fe/H]} = 0$ and most stars agreeing within $\sim 0.2$ dex, on order with our external accuracy as determined in \autoref{sec:GCS}.  Our large uncertainties with respect to age are evident in \autoref{fig:deltas}, with some Myr stars being matched to Gyr counterparts.  This is likely due to pairs containing members along the giant branch and main sequence.  In \autoref{fig:CMDs}, many of the youngest stars can be found along the RGB, which suggests those ages are not trustworthy.  These stars are being assigned ages appropriate for pre-main sequence (PMS) stars, which also affects their isochronal distance estimates, since RGB and PMS stars have much different luminosities. We confirmed this by comparing the parallactic and isochronal distances, and many RGB stars with erroneous young ages have large differences (> 50 pc) in their distance estimates.

\begin{figure}\label{fig:differences}
\plottwo{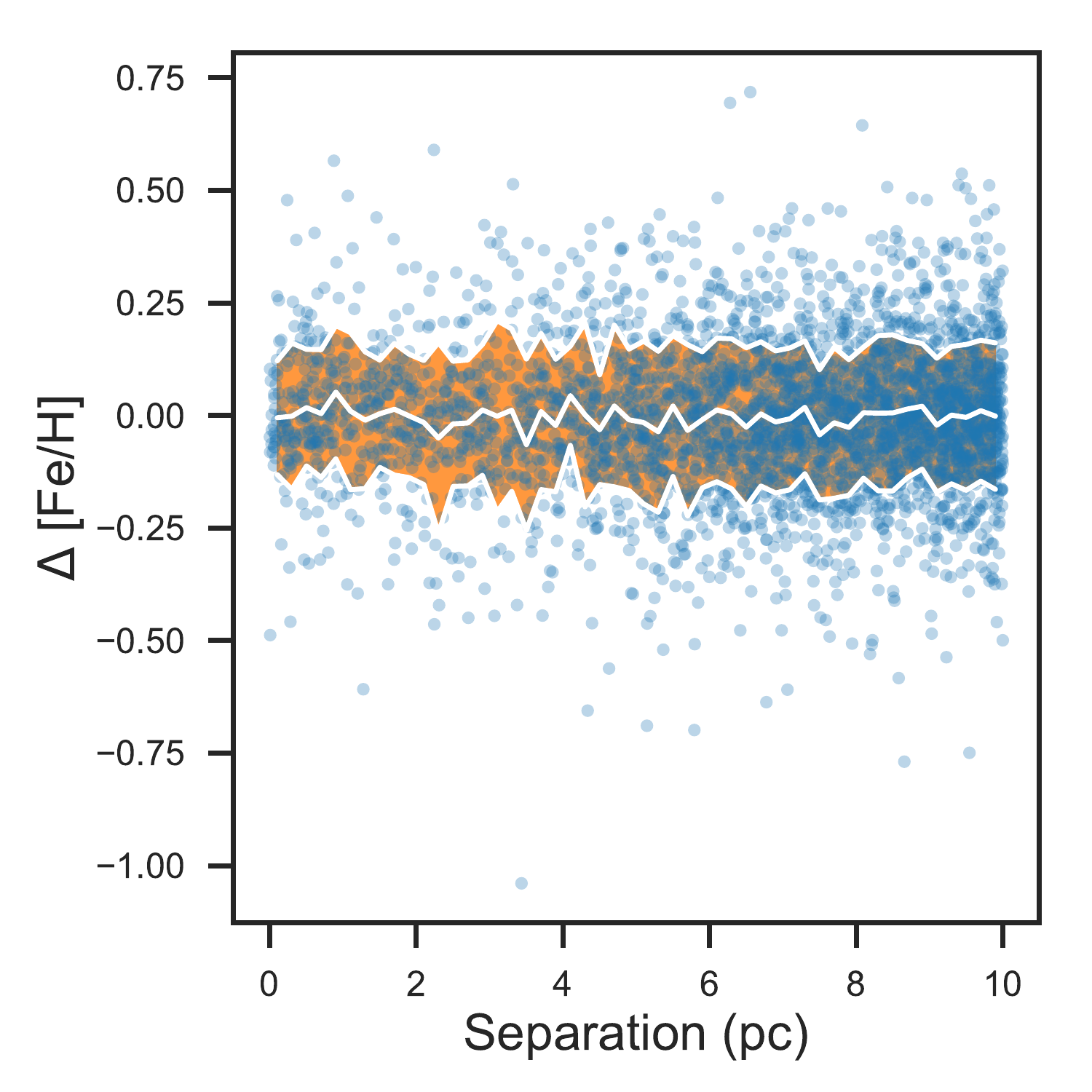}{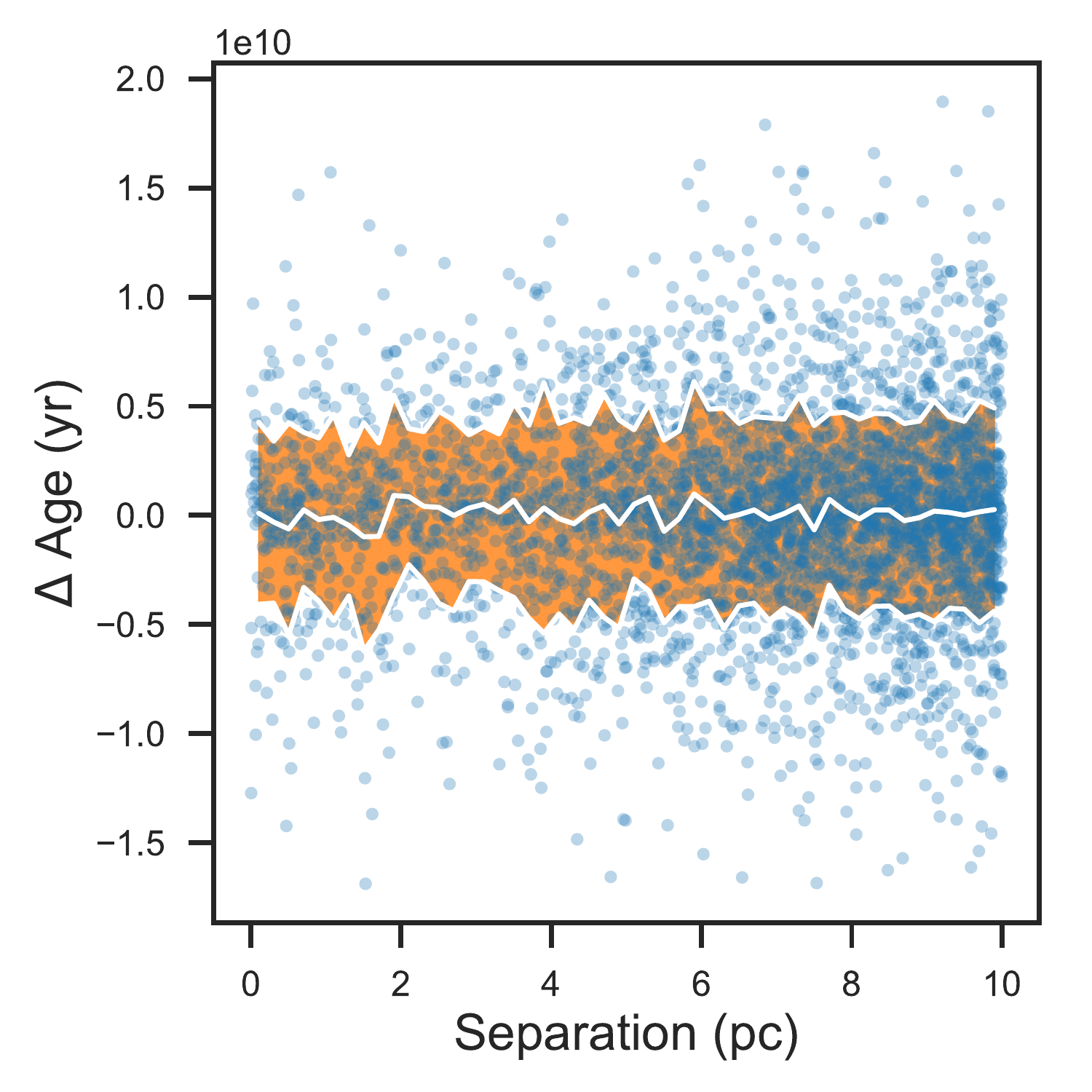}
\caption{Left panel: Difference in estimated [Fe/H] vs. physical separation in pc.  Right Panel: Difference in estimated age vs. physical separation in pc. Each filled circle represents a single binary system, and the means and standard deviations in 50 equally spaced bins are overplotted as the solid white line and filled orange area.}
\end{figure}

We also examined the differences in age and metallicity as a function of the separation of the individual stars. In Figure \autoref{fig:differences}, we show the differences as a function of separation, as well as the mean and standard deviation among 50 bins.  There are no clear correlations with physical separation, which suggests that the pairs with large separations may be bona fide binaries.  We also examined the distributions in age and metallicity differences in randomly associated pairs from our binary sample.  For both age and metallicity, the standard deviation of differences was larger for the randomly associated stars as compared to our original sample.

Despite the uncertainties in $[Fe/H]$ and age, we searched for binaries with identical members in terms of mass, age and composition.  To select these twins, we enforced that the median estimates for both stars needed to match within 0.015 in mass, $[Fe/H]$ and log age.  Three pairs of twins were identified, and summarized in \autoref{table:twins}.

Finally, we examined the ten pairs with the largest physical and projected separations.  Their properties are summarized in \autoref{table:large_sep} and \autoref{table:large_proj_sep}.  The pairs in \autoref{table:large_sep} are presented in order of increasing separation, but they all have separations $\sim$ 10 pc, while the projected distances vary.  As discussed in \cite{2017AJ....153..257O}, the co-moving stars separated by the largest distances are most prone to false positives.  Fundamental property estimates can test this, as false positives are unlikely to have the same metallicity and age.  For the small sample in \autoref{table:large_sep}, that is not an issue, as many have metallicity estimates that agree within their uncertainties.  Given the difficulty in photometrically estimating ages, coevality as determined by isochrone fitting cannot reliably rule out false positives.  We also note that at least two pairs in \autoref{table:large_sep} are main sequence stars paired with a red giant star.

\begin{deluxetable*}
{lrrrrrrrrc}
\caption{Stellar Twins}\label{table:twins}
\tablehead{
\colhead{TGAS Source ID} &
\colhead{R.A. (deg)} &
\colhead{Dec. (deg)} &
\colhead{Mass ($M_\odot$)} &
\colhead{Radius ($R_\odot$)}&
\colhead{[Fe/H] (dex)}&
\colhead{log(age (yr))}&
\colhead{Distance (pc)}&
\colhead{$A_v$ (mag)} & 
\colhead{Sep. (pc)}
}
\startdata
3152809288677876992 & 106.1067 &   5.6167 & $1.30^{+0.05}_{-0.05}$ & $1.46^{+0.08}_{-0.09}$ & $0.03^{+0.12}_{-0.08}$ & $9.28^{+0.22}_{-0.14}$ & $140.3^{+ 7.5}_{- 7.6}$ & $0.09^{+0.06}_{-0.09}$ & 8.6 \\
3128405559378754560 & 104.8429 &   4.1459 & $1.29^{+0.06}_{-0.07}$ & $1.42^{+0.07}_{-0.08}$ & $0.04^{+0.16}_{-0.11}$ & $9.27^{+0.33}_{-0.19}$ & $130.9^{+ 5.1}_{- 5.2}$ & $0.16^{+0.11}_{-0.16}$ &  8.6 \\
\hline
5910699048908083840 & 262.1762 & -62.4781 & $0.82^{+0.05}_{-0.06}$ & $0.86^{+0.03}_{-0.03}$ & $0.02^{+0.14}_{-0.15}$ & $10.06^{+0.35}_{-0.18}$ & $132.6^{+ 4.1}_{- 4.3}$ & $0.04^{+0.03}_{-0.02}$ & 8.1 \\
5816868066622162816 & 257.0792 & -64.2002 & $0.82^{+0.04}_{-0.05}$ & $0.85^{+0.03}_{-0.03}$ & $0.03^{+0.11}_{-0.13}$ & $10.07^{+0.36}_{-0.18}$ & $136.1^{+ 4.7}_{- 5.2}$ & $0.04^{+0.03}_{-0.04}$ &  8.1 \\
\hline
5860448893617125120 & 182.6068 & -66.5302 & $1.04^{+0.07}_{-0.07}$ & $1.10^{+0.06}_{-0.07}$ & $-0.06^{+0.16}_{-0.15}$ & $9.59^{+0.33}_{-0.24}$ & $183.8^{+10.4}_{-11.9}$ & $0.21^{+0.14}_{-0.19}$ & 9.5 \\
5236357160153303808 & 175.1802 & -66.5465 & $1.05^{+0.06}_{-0.06}$ & $1.12^{+0.06}_{-0.07}$ & $-0.07^{+0.13}_{-0.16}$ & $9.58^{+0.23}_{-0.19}$ & $182.7^{+10.0}_{-11.2}$ & $0.28^{+0.16}_{-0.16}$ &  9.5 \\
\enddata
\end{deluxetable*}

\begin{deluxetable*}
{lrrrrrrrrc}
\caption{Pairs with largest physical separations}\label{table:large_sep}
\tablehead{
\colhead{TGAS Source ID} &
\colhead{R.A. (deg)} &
\colhead{Dec. (deg)} &
\colhead{Mass ($M_\odot$)} &
\colhead{Radius ($R_\odot$)}&
\colhead{[Fe/H] (dex)}&
\colhead{log(age (yr))}&
\colhead{Distance (pc)}&
\colhead{$A_v$ (mag)} & 
\colhead{Sep. (pc)}
}
\startdata
1620697834607325056 & 222.4335 &  64.2362 & $0.81^{+0.01}_{-0.02}$ & $0.91^{+0.03}_{-0.03}$ & $-0.10^{+0.06}_{-0.06}$ & $10.12^{+0.02}_{-0.02}$ & $216.5^{+12.5}_{-12.8}$ & $0.01^{+0.00}_{-0.00}$ & 9.99 \\
1668177735991697536 & 216.1852 &  64.6618 & $1.13^{+0.03}_{-0.04}$ & $1.25^{+0.04}_{-0.05}$ & $-0.01^{+0.02}_{-0.00}$ & $9.53^{+0.15}_{-0.12}$ & $203.3^{+ 4.6}_{- 9.7}$ & $0.01^{+0.01}_{-0.00}$ &  9.99 \\
\hline
5367018414717465728 & 159.5562 & -44.4045 & $0.96^{+0.09}_{-0.01}$ & $0.88^{+0.03}_{-0.01}$ & $0.13^{+0.18}_{-0.06}$ & $9.36^{+0.06}_{-0.36}$ & $232.1^{+ 8.7}_{- 4.7}$ & $0.04^{+0.00}_{-0.01}$ & 9.99 \\
5366480031973428480 & 159.4605 & -46.2679 & $2.39^{+0.00}_{-0.00}$ & $10.55^{+0.35}_{-0.56}$ & $0.21^{+0.03}_{-0.02}$ & $8.92^{+0.02}_{-0.02}$ & $262.1^{+ 8.4}_{-12.6}$ & $0.09^{+0.03}_{-0.05}$ &  9.99 \\
\hline
2940892681712270464 &  93.1345 & -21.4982 & $1.71^{+0.10}_{-0.12}$ & $1.65^{+0.13}_{-0.15}$ & $-0.10^{+0.13}_{-0.16}$ & $8.68^{+0.21}_{-0.18}$ & $236.4^{+18.0}_{-22.5}$ & $0.04^{+0.02}_{-0.02}$ & 9.99 \\
2913756253703212416 &  93.0784 & -22.4562 & $0.83^{+0.04}_{-0.04}$ & $0.88^{+0.05}_{-0.04}$ & $0.07^{+0.16}_{-0.11}$ & $10.09^{+0.19}_{-0.13}$ & $244.3^{+14.5}_{-15.7}$ & $0.02^{+0.02}_{-0.02}$ &  9.99 \\
\hline
5321384833872077056 & 126.5134 & -52.6093 & $1.17^{+0.02}_{-0.01}$ & $1.38^{+0.04}_{-0.04}$ & $0.06^{+0.05}_{-0.06}$ & $9.63^{+0.07}_{-0.05}$ & $244.7^{+ 9.7}_{- 9.0}$ & $0.14^{+0.09}_{-0.14}$ & 9.99 \\
5319637916053876224 & 125.1789 & -54.4705 & $1.07^{+0.05}_{-0.04}$ & $1.07^{+0.06}_{-0.06}$ & $-0.05^{+0.09}_{-0.11}$ & $9.36^{+0.23}_{-0.19}$ & $270.4^{+20.1}_{-20.0}$ & $0.17^{+0.09}_{-0.08}$ &  9.99 \\
\hline
6519199260800456832 & 340.1186 & -46.4696 & $1.00^{+0.04}_{-0.02}$ & $0.92^{+0.03}_{-0.02}$ & $0.04^{+0.14}_{-0.07}$ & $9.14^{+0.30}_{-0.25}$ & $118.4^{+ 4.3}_{- 3.6}$ & $0.00^{+0.00}_{-0.00}$ & 9.99 \\
6517725709061049088 & 338.6571 & -47.5968 & $1.22^{+0.01}_{-0.01}$ & $1.73^{+0.11}_{-0.17}$ & $0.14^{+0.07}_{-0.05}$ & $9.67^{+0.04}_{-0.02}$ & $109.7^{+ 4.3}_{- 7.5}$ & $0.01^{+0.00}_{-0.00}$ &  9.99 \\
\hline
4440058987840821248 & 247.2641 &   7.5839 & $2.44^{+0.00}_{-0.00}$ & $11.15^{+0.61}_{-0.92}$ & $0.35^{+0.06}_{-0.05}$ & $8.93^{+0.02}_{-0.02}$ & $379.7^{+12.4}_{-16.2}$ & $0.01^{+0.01}_{-0.02}$ & 10.00 \\
4452408565005453312 & 246.5307 &   8.7792 & $1.33^{+0.03}_{-0.01}$ & $1.94^{+0.21}_{-0.11}$ & $0.35^{+0.06}_{-0.08}$ & $9.63^{+0.02}_{-0.02}$ & $480.8^{+27.7}_{-16.7}$ & $0.03^{+0.02}_{-0.01}$ &  10.00 \\
\hline
5525278854240810624 & 131.1946 & -40.7620 & $1.37^{+0.08}_{-0.09}$ & $1.46^{+0.09}_{-0.10}$ & $0.05^{+0.18}_{-0.09}$ & $9.09^{+0.32}_{-0.24}$ & $249.2^{+14.9}_{-16.3}$ & $0.43^{+0.21}_{-0.35}$ & 10.00 \\
5529087253282496768 & 129.8090 & -38.8548 & $1.05^{+0.06}_{-0.06}$ & $1.01^{+0.05}_{-0.06}$ & $-0.09^{+0.18}_{-0.14}$ & $9.18^{+0.66}_{-0.37}$ & $248.6^{+14.6}_{-16.5}$ & $0.13^{+0.09}_{-0.17}$ &  10.00 \\
\hline
5778955187704376320 & 243.8480 & -77.8374 & $0.98^{+0.05}_{-0.06}$ & $1.09^{+0.05}_{-0.06}$ & $-0.01^{+0.11}_{-0.12}$ & $9.84^{+0.21}_{-0.15}$ & $196.1^{+ 9.1}_{-10.3}$ & $0.05^{+0.04}_{-0.04}$ & 10.00 \\
5780781132920397952 & 240.9641 & -76.3968 & $0.76^{+0.00}_{-0.00}$ & $1.12^{+0.11}_{-0.12}$ & $-0.33^{+0.07}_{-0.06}$ & $10.28^{+0.02}_{-0.02}$ & $200.0^{+18.8}_{-18.1}$ & $0.04^{+0.02}_{-0.01}$ &  10.00 \\
\hline
5600596397178150784 & 116.9132 & -28.7256 & $0.95^{+0.05}_{-0.05}$ & $0.91^{+0.03}_{-0.04}$ & $-0.16^{+0.18}_{-0.17}$ & $9.42^{+0.43}_{-0.29}$ & $94.8^{+ 2.6}_{- 2.6}$ & $0.11^{+0.08}_{-0.14}$ & 10.00 \\
5605692083818444032 & 110.5832 & -29.6616 & $0.72^{+0.03}_{-0.04}$ & $0.71^{+0.02}_{-0.02}$ & $-0.05^{+0.12}_{-0.12}$ & $10.01^{+0.43}_{-0.21}$ & $97.8^{+ 2.7}_{- 2.9}$ & $0.08^{+0.06}_{-0.08}$ &  10.00 \\
\hline
313880985995995904 &  17.8932 &  32.5219 & $0.70^{+0.03}_{-0.04}$ & $0.69^{+0.02}_{-0.02}$ & $-0.03^{+0.11}_{-0.14}$ & $10.01^{+0.46}_{-0.22}$ & $44.0^{+ 0.6}_{- 0.6}$ & $0.03^{+0.02}_{-0.02}$ & 10.00 \\
2808682524506091648 &  12.4815 &  26.9230 & $0.92^{+0.02}_{-0.02}$ & $0.86^{+0.01}_{-0.01}$ & $0.01^{+0.05}_{-0.06}$ & $9.40^{+0.24}_{-0.24}$ & $51.9^{+ 0.8}_{- 0.9}$ & $0.02^{+0.01}_{-0.01}$ &  10.00 \\
\enddata
\end{deluxetable*}

\begin{deluxetable*}
{lrrrrrrrrc}
\caption{Pairs with largest projected separations}\label{table:large_proj_sep}
\tablehead{
\colhead{TGAS Source ID} &
\colhead{R.A. (deg)} &
\colhead{Dec. (deg)} &
\colhead{Mass ($M_\odot$)} &
\colhead{Radius ($R_\odot$)}&
\colhead{[Fe/H] (dex)}&
\colhead{log(age (yr))}&
\colhead{Distance (pc) \tablenotemark{a}}&
\colhead{$A_v$ (mag)} & 
\colhead{Proj  Sep. (deg.)}
}
\startdata
1151205132596390400 & 135.6178 &  86.6559 & $0.91^{+0.03}_{-0.02}$ & $0.86^{+0.01}_{-0.02}$ & $0.02^{+0.02}_{-0.06}$ & $9.59^{+0.22}_{-0.25}$ & $50.0^{+ 0.7}_{- 0.8}$ & $0.05^{+0.04}_{-0.04}$ & 10.96 \\
552966731438640640 &  75.2248 &  77.7755 & $0.99^{+0.01}_{-0.02}$ & $1.27^{+0.03}_{-0.04}$ & $-0.16^{+0.06}_{-0.08}$ & $9.87^{+0.02}_{-0.02}$ & $50.8^{+ 0.8}_{- 0.8}$ & $0.05^{+0.03}_{-0.04}$ &  10.96 \\
\hline
6791843131915506176 & 308.6003 & -33.7682 & $0.91^{+0.03}_{-0.02}$ & $0.84^{+0.02}_{-0.02}$ & $0.17^{+0.12}_{-0.09}$ & $9.57^{+0.22}_{-0.20}$ & $44.3^{+ 0.7}_{- 0.7}$ & $0.01^{+0.01}_{-0.01}$ & 11.14 \\
6741888092418571776 & 295.3066 & -32.8739 & $0.89^{+0.02}_{-0.03}$ & $1.11^{+0.04}_{-0.04}$ & $-0.08^{+0.13}_{-0.10}$ & $10.07^{+0.02}_{-0.02}$ & $46.6^{+ 1.0}_{- 1.1}$ & $0.02^{+0.02}_{-0.06}$ &  11.14 \\
\hline
6507617417630891008 & 340.0169 & -52.0083 & $0.61^{+0.02}_{-0.03}$ & $0.59^{+0.02}_{-0.02}$ & $0.03^{+0.12}_{-0.14}$ & $10.02^{+0.48}_{-0.21}$ & $48.0^{+ 0.9}_{- 0.9}$ & $0.01^{+0.00}_{-0.00}$ & 11.20 \\
6466670333304011264 & 322.3107 & -50.3169 & $1.30^{+0.01}_{-0.01}$ & $1.30^{+0.02}_{-0.04}$ & $0.17^{+0.05}_{-0.08}$ & $8.94^{+0.14}_{-0.24}$ & $50.0^{+ 0.9}_{- 0.9}$ & $0.01^{+0.01}_{-0.01}$ &  11.20 \\
\hline
6456068086272371712 & 316.3829 & -59.1385 & $0.72^{+0.02}_{-0.01}$ & $0.68^{+0.01}_{-0.01}$ & $-0.18^{+0.06}_{-0.06}$ & $9.54^{+0.14}_{-0.22}$ & $32.8^{+ 0.3}_{- 0.3}$ & $0.01^{+0.00}_{-0.02}$ & 11.42 \\
6665685408263289600 & 299.0673 & -52.9715 & $0.63^{+0.02}_{-0.03}$ & $0.62^{+0.01}_{-0.02}$ & $-0.00^{+0.09}_{-0.13}$ & $10.12^{+0.35}_{-0.15}$ & $31.5^{+ 0.3}_{- 0.3}$ & $0.02^{+0.01}_{-0.02}$ &  11.42 \\
\hline
4526375460984071296 & 273.8265 &  18.5002 & $0.72^{+0.01}_{-0.01}$ & $0.66^{+0.01}_{-0.01}$ & $0.03^{+0.02}_{-0.04}$ & $8.63^{+0.03}_{-0.25}$ & $30.6^{+ 0.6}_{- 0.7}$ & $0.08^{+0.05}_{-0.04}$ & 11.53 \\
4590489976865508480 & 272.3394 &  29.9520 & $1.03^{+0.02}_{-0.02}$ & $0.98^{+0.02}_{-0.02}$ & $0.05^{+0.08}_{-0.06}$ & $9.27^{+0.27}_{-0.21}$ & $24.6^{+ 0.2}_{- 0.2}$ & $0.03^{+0.02}_{-0.02}$ &  11.53 \\
\hline
6654965685287948032 & 281.3434 & -51.4210 & $0.56^{+0.01}_{-0.03}$ & $0.54^{+0.01}_{-0.03}$ & $-0.17^{+0.14}_{-0.18}$ & $9.83^{+0.11}_{-0.17}$ & $33.9^{+ 0.6}_{- 0.5}$ & $0.02^{+0.01}_{-0.01}$ & 12.13 \\
6439391793415687168 & 280.3194 & -63.5343 & $0.77^{+0.03}_{-0.03}$ & $0.74^{+0.02}_{-0.02}$ & $-0.02^{+0.11}_{-0.06}$ & $9.87^{+0.30}_{-0.23}$ & $37.7^{+ 0.5}_{- 0.4}$ & $0.05^{+0.03}_{-0.02}$ &  12.13 \\
\hline
4292775144697480704 & 290.3652 &   4.2065 & $0.68^{+0.02}_{-0.02}$ & $0.64^{+0.01}_{-0.01}$ & $-0.14^{+0.10}_{-0.16}$ & $9.50^{+0.22}_{-0.32}$ & $38.3^{+ 0.6}_{- 0.5}$ & $0.53^{+0.30}_{-0.23}$ & 12.30 \\
4321170822755315968 & 289.6880 &  16.4883 & $1.27^{+0.05}_{-0.01}$ & $1.21^{+0.04}_{-0.01}$ & $-0.21^{+0.08}_{-0.03}$ & $8.09^{+0.07}_{-0.54}$ & $36.2^{+ 0.8}_{- 0.7}$ & $0.79^{+0.17}_{-0.14}$ &  12.30 \\
\hline
1490845580086869760 & 217.3920 &  39.7904 & $0.82^{+0.03}_{-0.03}$ & $0.78^{+0.02}_{-0.02}$ & $-0.07^{+0.11}_{-0.14}$ & $9.69^{+0.34}_{-0.22}$ & $42.4^{+ 0.6}_{- 0.6}$ & $0.00^{+0.00}_{-0.00}$ & 12.47 \\
1476485992587837440 & 201.2826 &  38.9224 & $0.76^{+0.01}_{-0.01}$ & $0.72^{+0.01}_{-0.00}$ & $0.02^{+0.02}_{-0.02}$ & $9.64^{+0.12}_{-0.24}$ & $42.5^{+ 0.4}_{- 0.5}$ & $0.00^{+0.00}_{-0.00}$ &  12.47 \\
\hline
1327719251850566656 & 248.2193 &  35.0751 & $0.91^{+0.02}_{-0.01}$ & $0.87^{+0.02}_{-0.01}$ & $0.09^{+0.07}_{-0.04}$ & $9.71^{+0.16}_{-0.10}$ & $34.7^{+ 0.3}_{- 0.4}$ & $0.01^{+0.00}_{-0.01}$ & 12.55 \\
1389639211242104704 & 234.0527 &  40.8321 & $0.60^{+0.01}_{-0.01}$ & $0.58^{+0.01}_{-0.01}$ & $-0.12^{+0.06}_{-0.09}$ & $9.79^{+0.20}_{-0.28}$ & $38.2^{+ 0.6}_{- 0.6}$ & $0.01^{+0.00}_{-0.01}$ &  12.55 \\
\hline
4753362798950483072 &  40.5352 & -46.5248 & $1.45^{+0.00}_{-0.01}$ & $2.48^{+0.10}_{-0.15}$ & $0.07^{+0.05}_{-0.07}$ & $9.43^{+0.02}_{-0.01}$ & $42.4^{+ 0.6}_{- 0.7}$ & $0.01^{+0.01}_{-0.00}$ & 13.17 \\
4960514947152086016 &  24.5134 & -40.2936 & $0.67^{+0.01}_{-0.01}$ & $0.63^{+0.01}_{-0.01}$ & $0.11^{+0.05}_{-0.05}$ & $9.34^{+0.20}_{-0.29}$ & $38.9^{+ 0.5}_{- 0.5}$ & $0.01^{+0.00}_{-0.00}$ &  13.17 \\
\enddata
\tablenotetext{a}{Distances are estimated from isochronal fits.}
\end{deluxetable*}

\subsubsection{Binary Mass Properties}
For each binary system, we calculated the separation (in pc and AU), the mass ratio, defined as the mass of the secondary divided by the mass of the primary star, the gravitational binding energy and the dissipative lifetime.  The binding energy was calculated using:
\begin{equation}
U = \frac{GM_1M_2}{R}
\end{equation}
where $G$ is the universal gravitational constant, $M_1$ and $M_2$ are the masses of the stars and $R$ is the physical separation between the two components.  The dissipative lifetime of the binaries was estimated using:
\begin{equation}
\tau = \frac{1.212 * (M_1+M_2)}{R}
\end{equation}
with $\tau$ in Gyr, $M_1$ and $M_2$ in solar masses and R in pc, from \cite{2010AJ....139.2566D}.

\begin{figure*}\label{fig:corner_mass}
\plotone{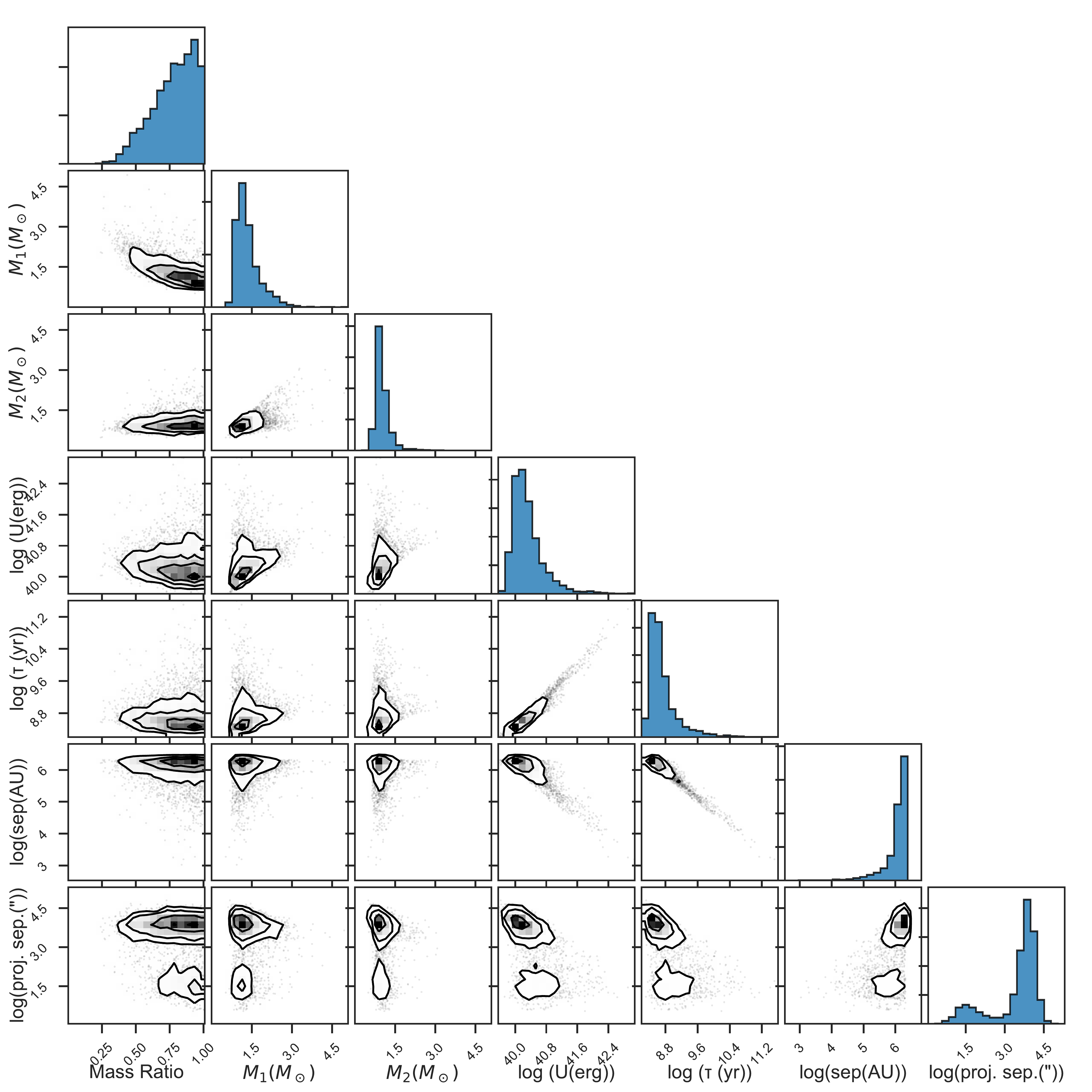}
\caption{A corner plot summary of the mass properties of binaries in our study.  We show the mass ratio, primary and secondary masses in solar masses, binding energy in ergs (log scale), binary lifetime in Gyr (log scale), physical separation in AU (log scale), and projected separation in arcseconds (log scale).  The most tightly bound, and longest lived systems are found close to each other.}
\end{figure*}

\begin{figure*}\label{fig:binding_energy}

\gridline{\fig{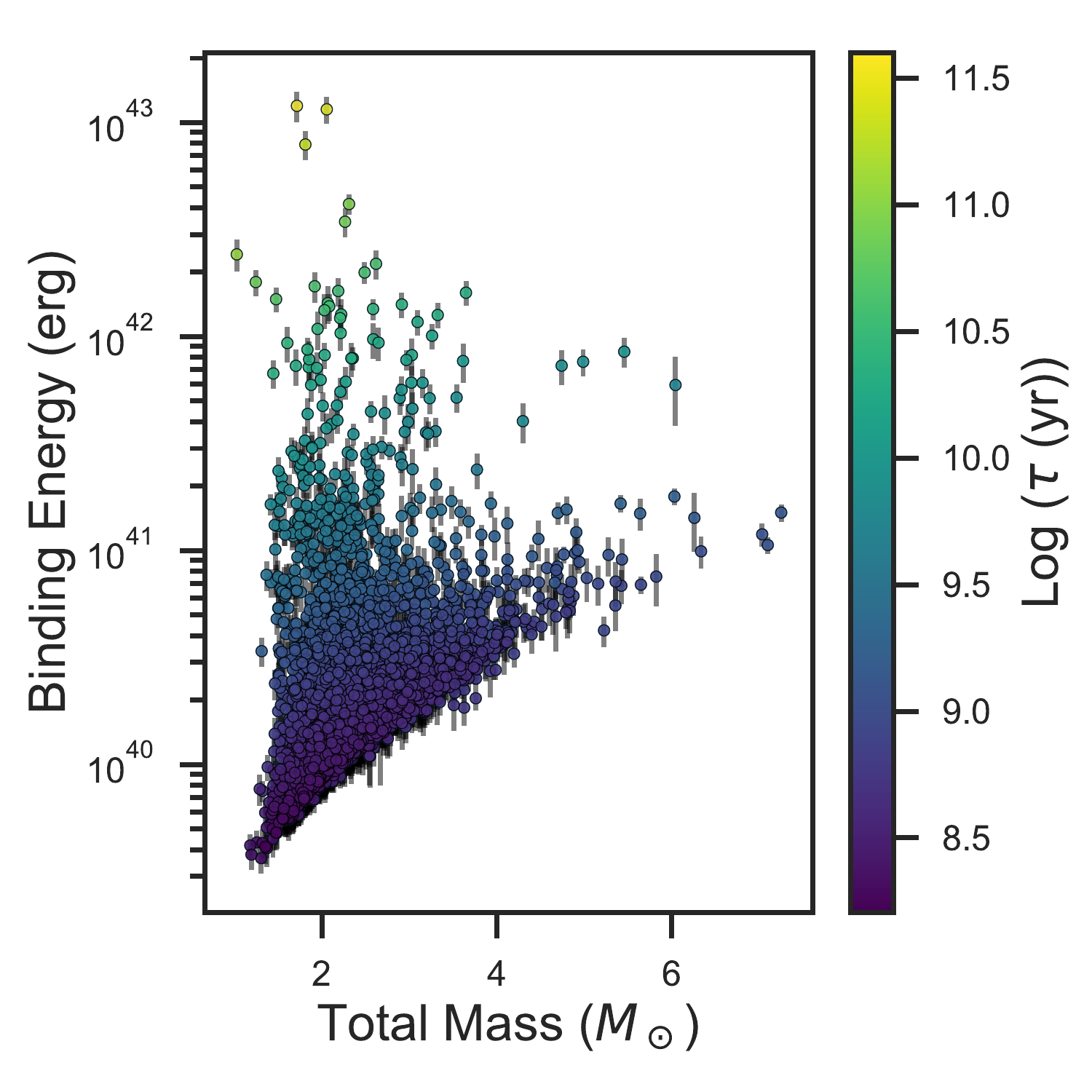}{0.5\textwidth}{}
          \fig{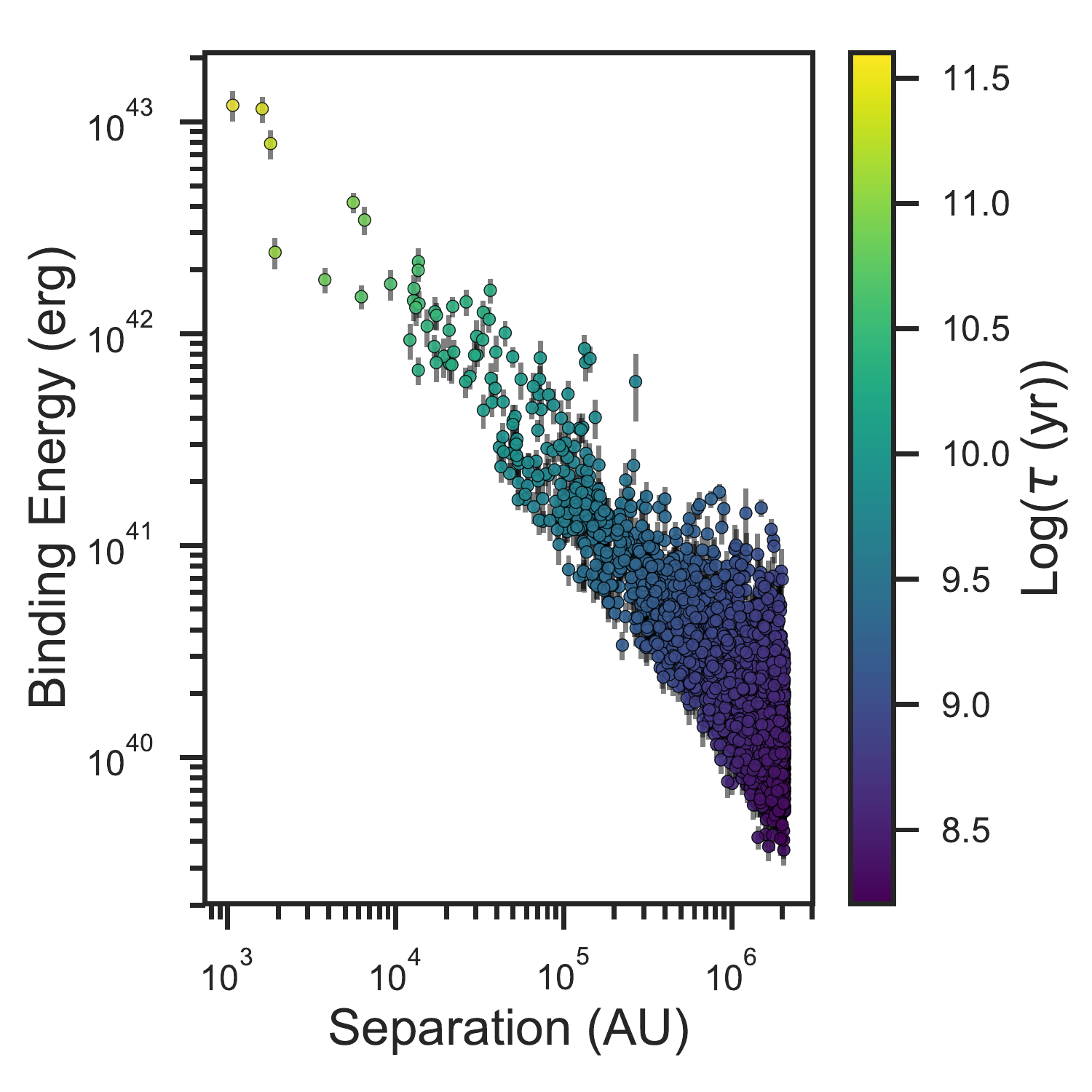}{0.5\textwidth}{}}
\gridline{\fig{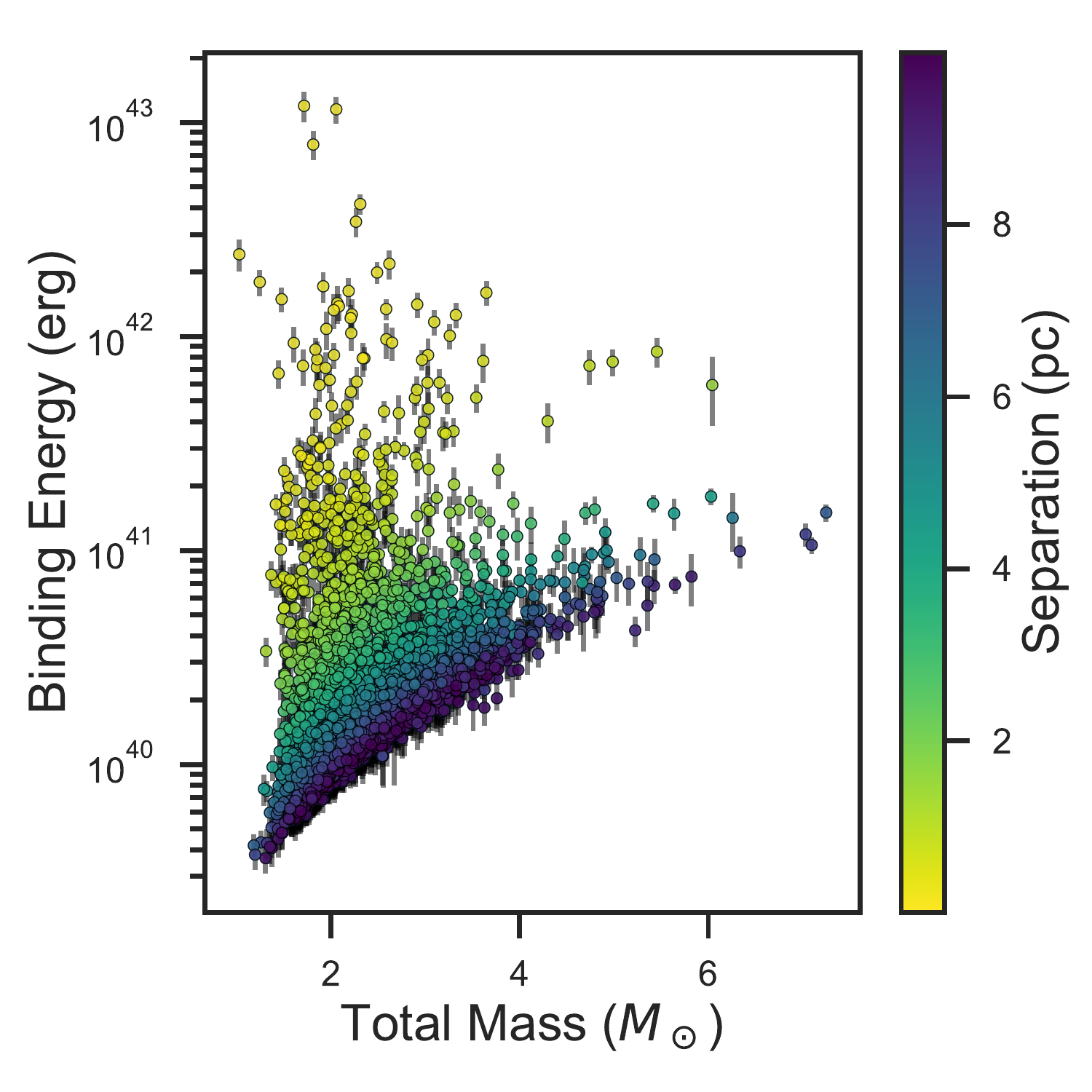}{0.5\textwidth}{}
          \fig{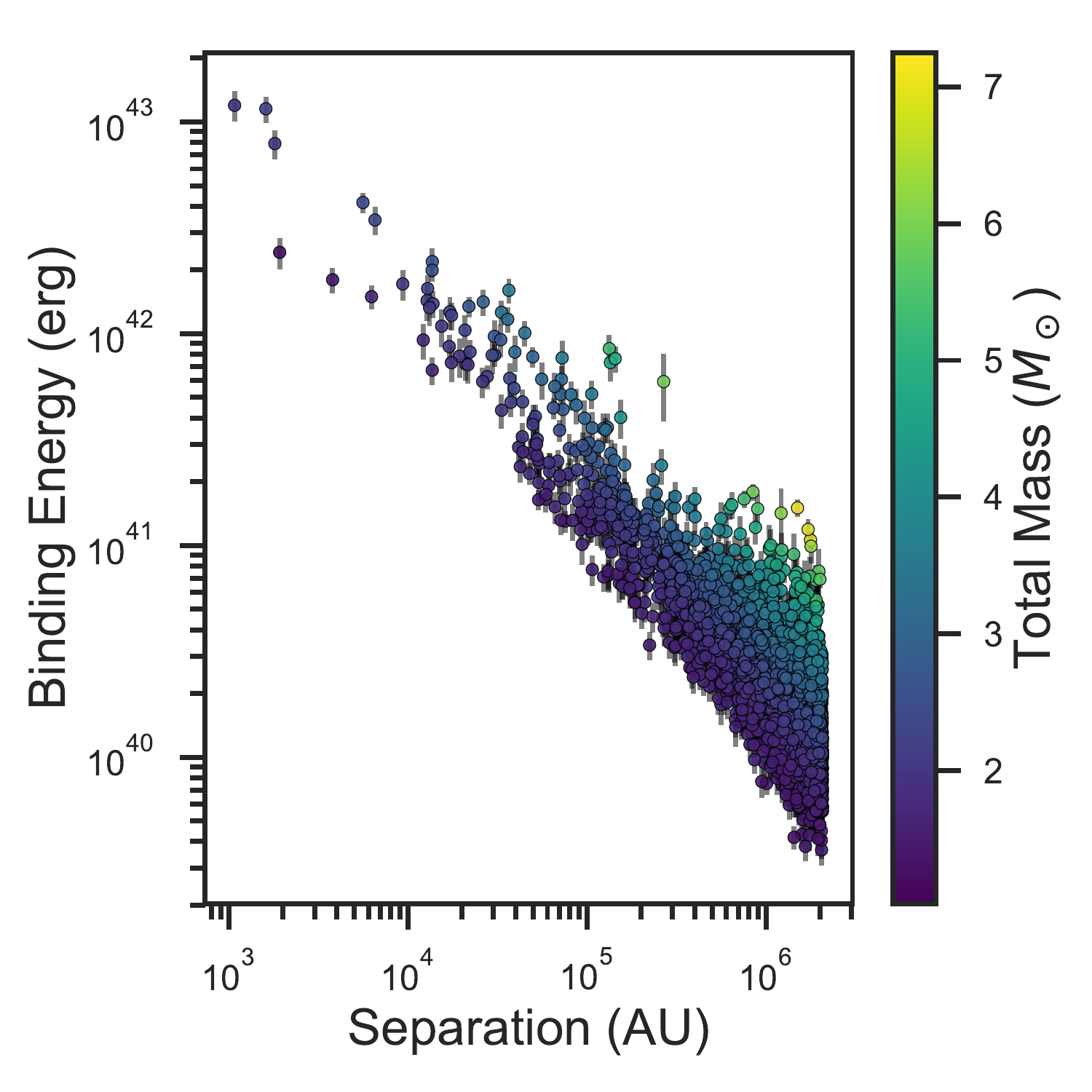}{0.5\textwidth}{}}

\caption{Left Panels: Binding Energy vs Total Mass for binaries in our sample.  Right Panels: Binding Energy vs Separation for the same binaries. The upper panels have points shaded by dissipation timescales, which approximates how long the binary should remain gravitationally bound.  The lower panels have points shaded by separation (in pc) and total mass (in $M_\odot$). The longest lived binaries in our sample are seen with small separations, and not necessarily large total masses. At the largest separations ($> 10^6 $ AU), there is a preference for larger total masses, suggesting that some of these widely separated pairs may be truly bound.  Overall, most of the sample consists of weakly bound systems ($U \sim 10^{40}$ ergs), suggesting they are easily disrupted.  These systems may have been more tightly bound in the past, but have since drifted apart due to gravitational interactions with neighboring stars or the Galactic tides.}
\end{figure*}

\begin{figure}\label{fig:proj_BE}
\plotone{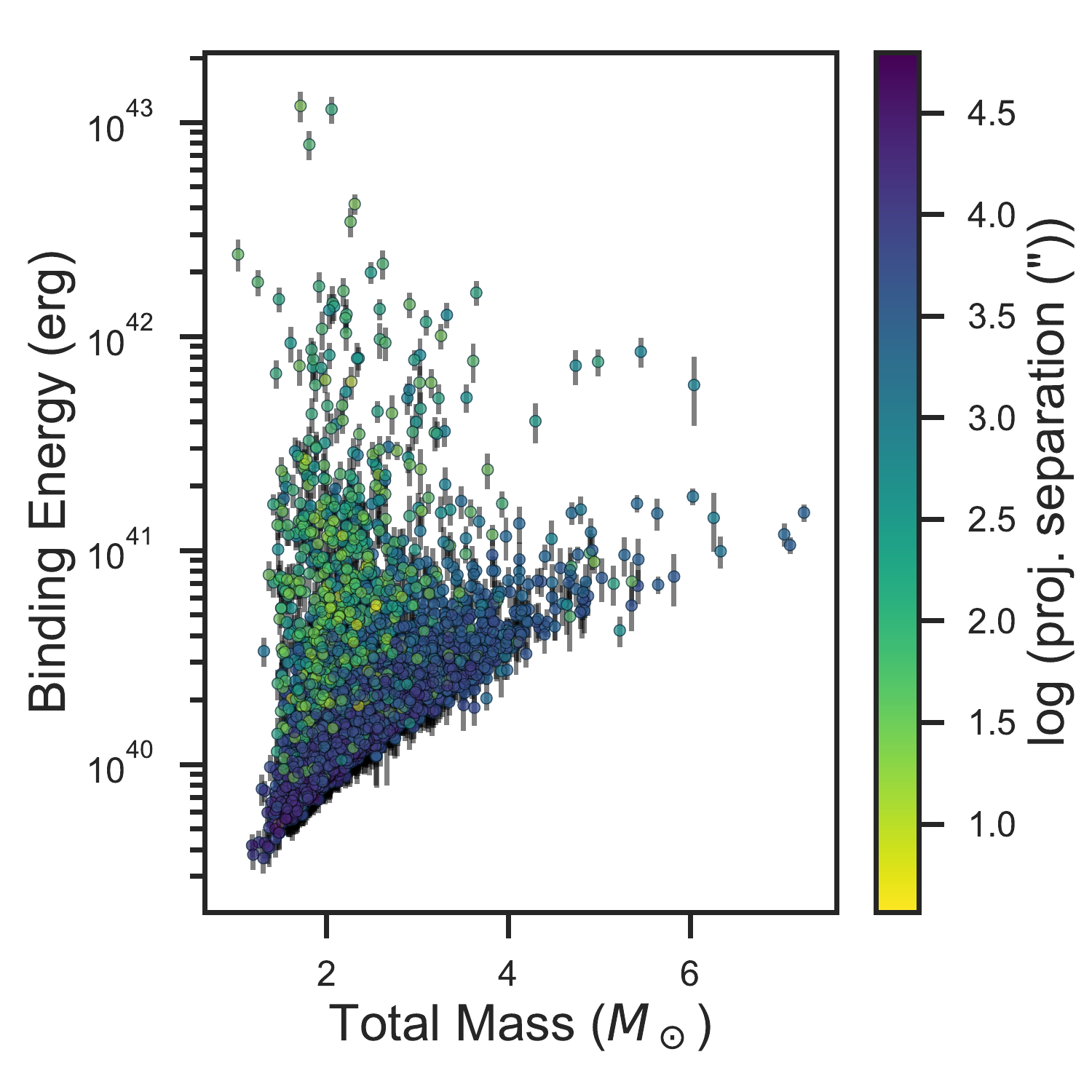}
\caption{Binding Energy vs. Total Mass, shaded by projected separation (log scale, in arcseconds). While the trends seen in \autoref{fig:binding_energy} are not as apparent, the pairs with the largest binding energies are also found at the smallest projected separations.}
\end{figure}

In \autoref{fig:corner_mass}, we show the summary corner plot of the mass properties of pairs in our sample.  Some trends are evident.  First, the mass ratio distribution rises towards values of unity.  The binding energy and lifetimes trend together, with the most tightly bound binaries having the largest lifetimes.  The separation between components has the largest influence on binding energy and lifetime, as it varies to a larger extent than the masses of the binary components.  This is also evident in \autoref{fig:binding_energy} which compares the binding energy to the separation between binary components and the total mass of the binary.  The median uncertainty in binding energy for the sample is $\sim 20\%$, with the uncertainty in the physical separation of the binary components being the largest factor.  Since the uncertainty in separation is dominated by the relative uncertainty in parallax, these uncertainties are small, due to the exquisite precision of $Gaia$. Tightly bound components, with large binding energies and long lifetimes, are only found at small separations.  This trend is also evident when the projected separation of the system is considered, as shown in \autoref{fig:proj_BE}. The projected separations do not track the binding energy as directly as the 3d separation, by definition, but there is a trend towards the closest stars (on the sky) also having large binding energies.
On the other hand, the total mass of the most tightly bound systems are not necessarily large, with many having total masses less than 2 solar masses.  We see that many stars in the sample have relatively low binding energies (and large separations) as noted in \cite{2017AJ....153..257O} and \cite{2017AJ....153..259O}.  These stars are likely formerly bound systems with similar space motions.

\begin{figure}\label{fig:mass_ratios}
\plotone{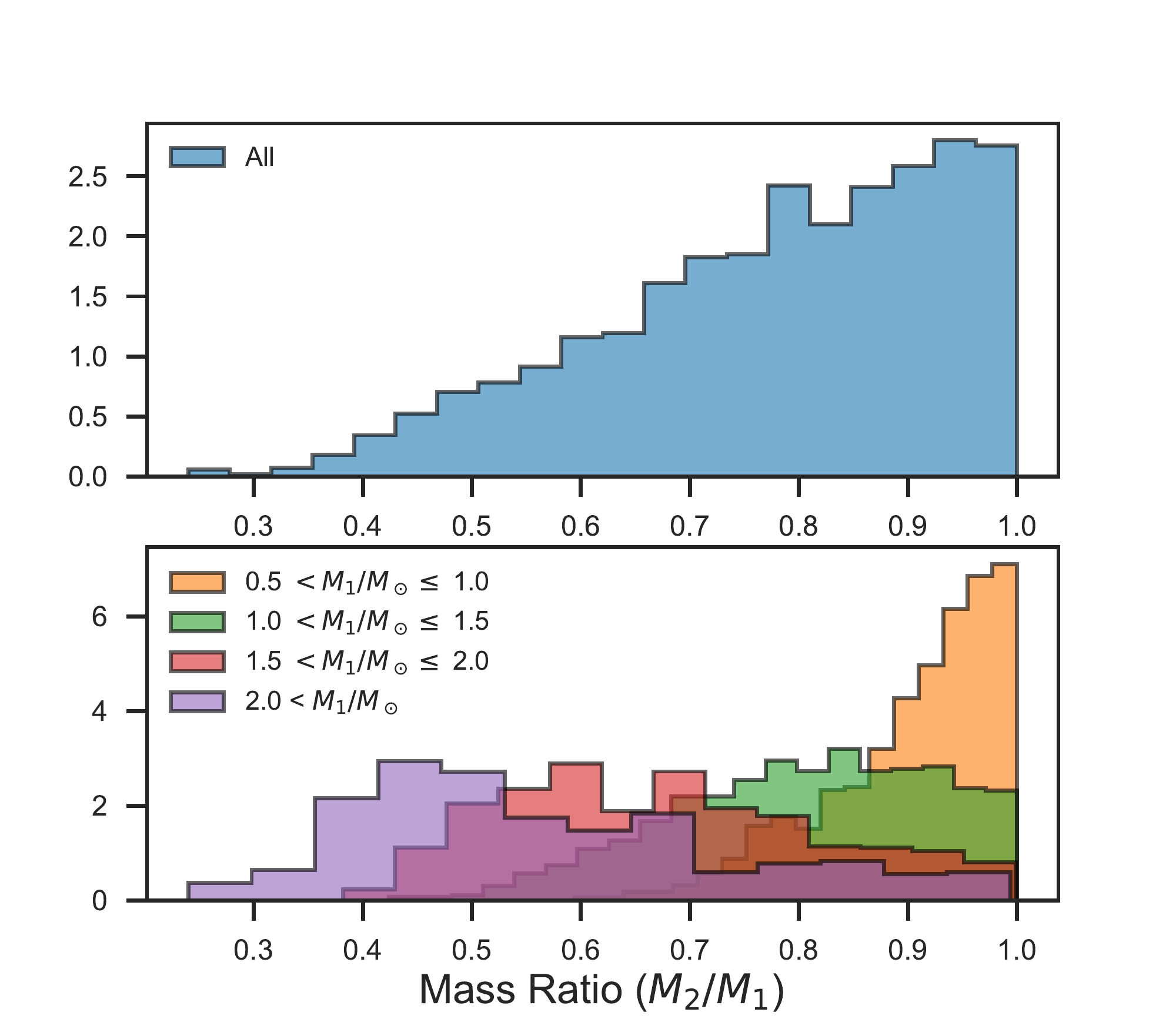}
\caption{Upper Panel: Mass ratio histogram for stars in our sample.  Lower Panel:  Mass ratio distribution for various bins in primary mass.  As the mass of the primary increases, the mass ratio distribution moves away from unity to increasingly flatter distributions, indicating that the secondary masses are drawn from the same distribution as the primary.}
\end{figure}

In \autoref{fig:mass_ratios}, we plot the overall mass ratio distribution, defined as the mass of the secondary divided by the mass of the primary component of each binary.  Overall, the mass ratio distribution grows as the ratio gets closer to unity.  In the lower panel, we divide the distribution in terms of the masses of the primary.  A clear trend towards flatter distributions arise as the mass of the primary increases.  This is partly due to the definition, as the lowest mass primary stars can only have companions that are relatively equal in terms of mass, while as the primary's mass increases, those stars can be paired with secondaries of a variety of masses.  

\section{Conclusions}\label{sec:conclusions}
We analyzed a sample of 9,995 co-moving stars identified in $Gaia$-TGAS by \cite{2017AJ....153..257O} and reorganized by \cite{faherty2017}.  Our analysis used isochrone fitting incorporating $G,J,H,K$ and $W1$ observed photometry.  Our results demonstrated robust estimates of stellar masses, as verified by comparisons to other analyses of the same stars, cluster properties and computation of the PDMF. We report fundamental parameters for all stars in the sample, and examined the binary properties of pairs in the system.  Many pairs in the catalog are weakly bound, sharing binding energies comparable to Neptune and the Sun \citep{2010AJ....139.2566D}.  The dominant property of a pair's binding energy and lifetime is its physical separation.  

This catalog, derived from the exquisite astrometry of $Gaia$ has revealed a large population of loosely bound systems, which will be ideal for large spectroscopic follow-up \citep[i.e.,][]{2017arXiv170903532P}.  Future samples of co-moving stars in $Gaia$ will likely be dominated by loosely bound systems, making spectroscopic observations critically important.  Spectra readily reveals the radial velocity of a star, allowing for the direct comparison of a star's 3d velocity vector to any companion.  Furthermore, spectra can be used to precisely estimate stellar chemical composition, which is uncertain with photometric techniques.  However, to address the significant age uncertainties, other techniques, like gyrochronology \citep[i.e.,]{2014ApJ...795..161D} or astroseismology \citep{2014ApJS..210....1C} are likely required.

\acknowledgments
The authors thank Semyeong Oh for fruitful discussions and openly hosting her work and catalogs. We thank the anonymous referee for suggestions that improved the clarity of this work.  We thank the many contributors to open source software that made this work possible.  This project was developed in part at the 2016 NYC Gaia Sprint, hosted by the Center for Computational Astrophysics at the Simons Foundation in New York City. 

JJB acknowledges the funding of Rider University's summer research grant. 

JJB dedicates this manuscript to the memory of his mother, Ginny Bochanski, who supported his first astronomical inquiries.

This research made use of Astropy, a community-developed core Python package for Astronomy \citep{2013A&A...558A..33A}.

This work has made use of data from the European Space Agency (ESA)
mission {\it Gaia} (\url{https://www.cosmos.esa.int/gaia}), processed by
the {\it Gaia} Data Processing and Analysis Consortium (DPAC,
\url{https://www.cosmos.esa.int/web/gaia/dpac/consortium}). Funding
for the DPAC has been provided by national institutions, in particular
the institutions participating in the {\it Gaia} Multilateral Agreement. 

This publication makes use of data products from the Wide-field Infrared Survey Explorer, which is a joint project of the University of California, Los Angeles, and the Jet Propulsion Laboratory/California Institute of Technology, funded by the National Aeronautics and Space Administration. 

This publication makes use of data products from the Two Micron All Sky Survey, which is a joint project of the University of Massachusetts and the Infrared Processing and Analysis Center/California Institute of Technology, funded by the National Aeronautics and Space Administration and the National Science Foundation.

\facilities{Gaia, CTIO:2MASS, FLWO:2MASS, WISE} 
\software{Numpy \citep{5725236}, jupyter \citep{jupyter}, scipy \citep{scipy}, isochrones \citep{2015ascl.soft03010M}, astroML \citep{2014ascl.soft07018V,astroML,astroMLText}, emcee \citep{2013PASP..125..306F}, corner \citep{2016JOSS.2016...24F}, BANYAN \citep{2018arXiv180109051G}, Matplotlib \citep{Hunter:2007}, Topcat \citep{2005ASPC..347...29T}}

\bibliographystyle{yahapj}
\bibliography{references}


\end{document}